\documentclass[11pt]{article}

\usepackage[a4paper, margin=1in]{geometry}
\usepackage{setspace}
\usepackage{comment}
\usepackage{appendix}

\usepackage{amsmath, amssymb, amsfonts}
\usepackage{booktabs}
\usepackage{array}
\usepackage{multirow}
\usepackage{makecell}
\usepackage{longtable}
\usepackage{threeparttable}
\usepackage{lscape}
\usepackage{pdflscape}
\usepackage{float}
\usepackage{graphicx}
\usepackage{subcaption}
\usepackage{xcolor}
\usepackage{natbib}

\usepackage{xr-hyper}
\usepackage{hyperref}

\hypersetup{
    colorlinks = true,
    linkcolor = blue,
    citecolor = blue,
    urlcolor = blue
}

\defcitealias{RCC_ENG}{RCS England, n.d.}
\usepackage{url}

\usepackage{cleveref}

\usepackage{multirow}
\usepackage[normalem]{ulem}
\useunder{\uline}{\ul}{} 

\hypersetup{
    colorlinks = true,
    linkcolor = blue,
    citecolor = blue,
    urlcolor = blue
}

\begin{document}

\begin{center}
{\Large\bfseries High-Volume, Low-Complexity Surgical Hubs in England:\\
Can They Improve Physician Productivity?}

\vspace{0.5cm}

Zecharias Anteneh$^{1,\ast}$, Adriana Castelli$^{1}$, Peter Sivey$^{1}$, Andrew Street$^{2}$, Jinglin Wen$^{1}$, Joy Adamson$^{3}$

\vspace{0.5cm}

$^{1}$ Centre for Health Economics, University of York, United Kingdom \\
$^{2}$ Department of Health Policy, London School of Economics and Political Science, United Kingdom \\
$^{3}$ Department of Health Sciences, University of York, United Kingdom \\[0.3cm]
$^{\ast}$ Corresponding author: \href{mailto:zecharias.anteneh@york.ac.uk}{zecharias.anteneh@york.ac.uk}
\end{center}
\begin{abstract}

Whether organisational separation of elective and emergency care improves physician productivity remains an open question. Most existing evidence relies on cross-sectional comparisons or volume-based outcomes that cannot isolate efficiency gains from input expansion or provider selection. In contrast, this paper exploits the staggered rollout of NHS England's surgical hub programme, a set of dedicated ring-fenced elective facilities introduced as part of its Elective Recovery Plan, to provide causal estimates of the effect of elective-emergency separation on physician productivity. Using this rollout to implement a heterogeneity-robust difference-in-differences design, we estimate the impact of hub adoption on physician productivity measured as cost-weighted elective output per unit of salary-weighted physician input. We find that hub adoption increases physician productivity by 14.5\% relative to the no-hub counterfactual, accompanied by a 10-day (7.6\%) reduction in average patient waiting times. The size of these gains depends on how completely elective care is insulated from emergency pathways. Standalone hubs situated on dedicated elective-only sites achieve relatively larger productivity gains than integrated hubs located within main acute hospital sites. Moreover, operating multiple hubs yields gains more than double the overall average, whereas a single hub shows no statistically significant effect. 
For health systems seeking to raise physician productivity and address elective backlogs, these findings suggest that how separation is implemented, not simply whether it is adopted, shapes the productivity gains it delivers.


\end{abstract}

\clearpage
\section{Introduction}

The productivity of skilled labour in public services is shaped not only by the capabilities of workers, but by the organisational environments in which they operate \citep{syverson2011determines, bartel2014human}. In hospital settings, this problem takes a specific and economically important form: senior physicians\footnote{In the NHS, senior hospital physicians are called hospital consultants. These are medical doctors who have completed specialist training and hold a substantive appointment to lead clinical care in a given specialty. The term is equivalent to attending physician in the United States, specialist physician in Australia and Canada, and \textit{Médecin spécialiste} or \textit{Facharzt} in France and Germany, respectively. Hospital consultants are the highest grade of hospital medical staff and the primary clinical labour input in elective surgical pathways.} perform elective procedures, manage emergency admissions and share the same wards, theatres, and scheduling systems. Emergency demand is inherently unpredictable, whereas elective care is planned. When both compete for the same inputs, the unpredictable crowds out the plannable, and productive capacity is systematically under-utilised \citep{johar2013emergency, aaserud_elective_2001, nasr_impact_2004}. Whether organisational separation of these functions can improve productivity is therefore an important question for health systems that rely on mixed emergency and elective provision but there is little causal evidence that if does so. 

This paper provides such evidence by exploiting the staggered rollout of surgical hubs across English \textit{National Health Service} (NHS) hospitals between April 2020 and March 2024. Surgical hubs (hereafter hubs) are ring-fenced elective facilities, physically and operationally separated from emergency services, designed to deliver High-Volume Low-Complexity (HVLC) procedures, including hip and knee replacements, cataract surgery, and similar standardised operations. These procedures are scheduled in advance but, in conventional hospital settings, are often exposed to emergency disruption, generating frequent cancellations and systematic underutilisation of planned theatre time. By isolating elective HVLC activity from emergency pressures, hubs aim to stabilise surgical schedules and enable more continuous use of physician time \citep{briggs2022optimising, co_impact_2025}.

Using monthly administrative data from Hospital Episode Statistics Admitted Patient Care linked to workforce and payroll data and unit cost information, and applying the heterogeneity-robust difference-in-differences (DiD) estimator of \citet{borusyak_revisiting_2024}, we estimate the causal effect of hub adoption on physician productivity by comparing each treated hospital's observed productivity to the counterfactual it would have achieved in the absence of a hub. 

We find that hub adoption increases physician productivity by approximately 14.5\% relative to the no-hub counterfactual. The dynamics of this effect are informative. At opening, physician productivity rises immediately, dips briefly in months one and two as scheduling protocols bed in, then recovers and grows steadily from month three, becoming consistently significant from month five and crossing \pounds0.50 worth of additional output per \pounds1 worth of physician input by month eight. 

The magnitude varies by type of hub. Standalone hubs, which operate on fully elective‑only sites, generate slightly larger gains than integrated hubs attached to the main hospital and much larger ring‑fenced hubs which by design located within it. The extent of adoption amplifies the effect: hospitals operating two or more hubs achieve gains more than double the overall average effect, while single-hub adoption yields effects statistically indistinguishable from zero. Single-specialty hubs generate larger gains than multi-specialty hub configurations (\pounds0.79 versus \pounds0.37 worth of additional output). These results point to a threshold effect in organisational design in which meaningful productivity gains require sufficiently complete separation and scale to insulate elective care from emergency disruption. 

Mechanism analysis provides suggestive evidence that part of the productivity gain, 17\%, is associated with a compositional shift toward HVLC activity, with hub adoption increasing the share of specialty activity accounted for by elective HVLC procedures; because this share is an outcome of adoption itself, we interpret the decomposition as descriptive rather than a causal mediation estimate. Hub adoption also reduces average patient waiting times by approximately 10 days, 7.6\% lower than expected under the no-hub counterfactual, consistent with the productivity gain translating into faster access to treatment. Neither length of stay nor operating theatre counts show statistically significant change, but the theatre-count evidence is imprecise and hub adoption bundles organisational separation with programme capital and funding support, so we do not interpret the gains as arising purely from reorganisation of existing resources.

This paper contributes to the literature in several ways. 
Direct evidence on the productivity effects of the NHS surgical hub or similar programmes in different contexts is limited. Two recent studies, \citealt{co_impact_2025} and \citealt{wen2026effect}, examine the effects of hub on surgical volumes, length of stay, and waiting times. Both studies report increases in procedure counts. The increases in total procedure volumes may not distinguish whether activity rose because more inputs were deployed or because existing inputs became more productive. We therefore focus on physician labour productivity, the margin most directly targeted by the programme and the metric relevant for assessing returns to organisational redesign. 

The surgical hub programme shares its organisational rational with two earlier examples of dedicated elective facility. In England, the introduction of both public and private specialised treatment centres are the closest historical precedent. \citet{siciliani2013differences} find that specialised public and private treatment centres achieved 16.7\% and 40\% shorter lengths of stay for hip replacement than NHS hospitals, attributing the difference to efficiency. \citet{barlow_effect_2013} document a two-day length-of-stay reduction following the introduction of a ring-fenced orthopaedic ward. 

In the United States, Medicare-certified Ambulatory Surgery Centres (ASCs) are the functional analogue. \citet{munnich_returns_2018} instrument for ASCs' utilisation using Medicare payment rates in order to make causal claims that they lowered post-operative hospital admission rates. \citet{courtemanche_does_2010} find that ASC entry is associated with a 2--4\% reduction in surgical activity at nearby hospitals, while \citet{hollenbeck_ambulatory_2015} document a 7\% reduction in hospital-based outpatient surgery rates following ASC opening. \citet{trentman_outpatient_2010} document shorter perioperative intervals at ASCs relative to hospital outpatient settings. \citet{martinussen_day_2004} find that increased day surgery rates improved hospital-level technical efficiency in Norwegian hospitals, with effects larger in hospitals with greater budgets consistent with our dose-response finding that scale amplifies the gains from organisational separation of elective and emergency care.

The policy relevance extends beyond England. Mixed elective-emergency provision is the norm across advanced health systems, and the same challenge exists everywhere, balancing planned care with unpredictable emergency demand. In our setting, hubs were adopted administratively, insurers do not select patients, and a reasonable number of hospitals did not adopt hubs; this combination mitigates several selection concerns that complicate market‑based US comparisons and makes our setting comparatively well suited to test the effects of dedicated elective provision. We emphasise that the treatment we evaluate is a policy bundle: hub adoption combines organisational separation with targeted programme capital support. Our estimates therefore capture the overall return to the programme as implemented, which is the object relevant to health systems considering similar investments to address elective backlogs under fiscal constraint. 

The remainder of the paper is organised as follows. Section \ref{sec:inst} presents the institutional background and surgical hub programme. Section \ref{sec:data} describes the data. Section \ref{sec:framework} sets out the conceptual framework while section \ref{sec:empirics} presents the empirical strategy. Section \ref{sec:results} reports main results, event study dynamics, heterogeneity analyses and robustness checks. Section \ref{sec:mechanisms} considers mechanisms that might be driving the observed effects. Sections \ref{sec:discussion} and \ref{sec:conclusion} present the discussion and concludes.

\section{Institutional Background}
\label{sec:inst}

\subsection{The NHS Elective Waiting List and the Case for Organisational Reform}

The National Health Service (NHS) in England is a publicly funded system providing comprehensive medical care free at the point of use, financed through general taxation and administered by NHS England. Hospital care is delivered predominantly through NHS hospitals — public hospitals that provide both emergency and elective services within the same institutional structure. This joint provision is central to the productivity problem this paper examines.

The elective waiting list stood at 4.4 million patients in March 2020, before the  COVID-19 pandemic \citep{nhsengland2020rtt}.  
The pandemic then produced a severe collapse in elective surgical activity: hospitals suspended planned procedures to manage emergency demand, protect staff capacity, and maintain infection control. The waiting list grew sharply as a result, exceeding 6 million patients waiting for treatment by mid-2021 and reaching a peak of 7.6 million in June 2023 \citep{nhsengland2023rtt, scobie2023strikes}. 

It has long been documented that emergency admissions routinely disrupt elective surgical lists: physicians are redeployed from planned procedures to manage acute cases, theatre time is reallocated at short notice, and recovery beds earmarked for elective patients are diverted to emergency admissions \citep{rcs2007separating, aaserud_elective_2001, addison2001separating}. Emergency demand is inherently unpredictable, while elective demand can be planned; but when both compete for the same theatres, wards, and physician time within the same hospital, it is the planned pathway that bears the disruption.  
It was in recognition of both this structural failure and in response to the post-pandemic backlog, that NHS England launched the surgical hub programme.

\subsection{The HVLC Surgical Hub Programme}

The High-Volume Low-Complexity (HVLC) surgical hub programme was formalised as a central component of the NHS Elective Recovery Plan in May 2021, though hub facilities had begun opening as early as April 2020 as part of earlier NHS England guidance on protecting elective activity during the pandemic. 
These early hubs share the same operational model as programme hubs but predate the formal programme structure. The programme established dedicated elective surgical facilities, referred to as hubs, and designed to physically and operationally separate high-volume standardised procedures from the acute hospital pathway, thereby insulating planned surgical lists from emergency disruption.

\subsection{HVLC Procedures: Definition and Justification}

The procedures targeted by the programme are HVLC procedures as defined by the Getting It Right First Time (GIRFT) programme, a national NHS initiative that develops clinical standards and procedure coding specifications for surgical pathways.\footnote{GIRFT coding specifications are publicly available at \url{https://gettingitrightfirsttime.co.uk}.}. These elective surgical procedures are characterised by high volume, low clinical complexity relative to emergency and tertiary surgery, and highly standardised pathways. It is precisely these features that make them well-suited to a focused, scheduling-protected production environment where physicians repeatedly perform a narrow set of procedures with dedicated support teams, rather than alternating between elective and emergency caseloads.

The surgical hub programme covers the following surgical specialties: Trauma and Orthopaedics ($T\&O$), General Surgery, Gynaecology, Ophthalmology, Urology, Ear Nose and Throat (ENT), and Spinal surgery.\footnote{Spinal surgery sits within the $T\&O$ clinical domain in the GIRFT coding framework but is treated as a distinct specialty in the operational structure of the hub's programme and in our data.} Representative HVLC procedures within each specialty include hip and knee replacement ($T\&O$), cholecystectomy (General Surgery), hysterectomy and laparoscopic procedures (Gynaecology), cataract extraction (Ophthalmology), transurethral resection and ureteroscopy (Urology), tonsillectomy and septoplasty (ENT), and spinal decompression and fusion (Spinal). These procedures collectively account for a substantial share of the elective backlog. They are also the operations most susceptible to cancellation during surges in emergency care, even though extended waiting times carry measurable adverse health effects on patients \citep{hodge2007consequences, hajat2002does}. The HVLC programme therefore targets procedures that are both highly amenable to delivery in dedicated elective facilities and particularly vulnerable to disruption and delay during periods of emergency pressure.

\subsection{Hub Organisational Models and Adoption}

The programme distinguishes between three organisational models, differing in the degree of physical and operational separation from the rest of the hospital \citep{tracey_elective_2024}. \textit{Ring-fenced hubs} operate as dedicated areas within acute hospital sites, featuring ring-fenced elective theatres within the main theatre complex, dedicated beds, and protected surgical teams (including senior doctors) to prevent diversion to emergency care. However, because they are integrated within the main hospital, they may share baseline site infrastructure, such as general ward capacity or, in limited cases, acute anaesthetic cover. \textit{Integrated hubs} are located within existing hospital sites, with all facilities, such as recovery areas and ward spaces, physically segregated from emergency pathways. While sharing the broader hospital footprint, these hubs operate via a separate entrance with strictly controlled access and ring-fenced surgical teams to ensure staff are not diverted to non-elective or emergency care \citepalias{RCC_ENG}.
\textit{Standalone hubs} are fully separate, dedicated facilities with no shared infrastructure with acute hospitals. Because the facility is entirely separate from acute care, there is no competition for operating theatres or staff including senior doctors to be used for urgent and emergency cases \citepalias{RCC_ENG}.

The degree of separation from emergency demand shocks increases from ring-fenced to integrated to standalone hubs. This separation gradient has a direct theoretical implication: the productivity gain from hub adoption should be increase in accordance with the degree to which there is separation between the emergency and elective pathways. We test this  prediction in our heterogeneity analysis in Section \ref{sec:results}.

The programme rolled out in a staggered fashion across NHS hospitals, with the pace of adoption determined by local capital readiness, estate availability, workforce configuration, and the timing of NHS England programme approval \citep{girft2021hubs}. These factors reflect implementation capacity rather than anticipated productivity outcomes, supporting the conditional exogeneity of adoption timing that our identification strategy requires. This institutional account is testable: the Supplementary Material (Section~\ref{S-sec:S1}, Table~\ref{S-tab:adoption_timing}) reports whether adoption, and adoption timing among adopters, are predicted by pre-programme hospital characteristics and pre-2020 productivity levels and trends. By the end of our study period in March 2024, 40 NHS hospitals had adopted at least one surgical hub serving one or more of the HVLC specialties.

\section{Data} 
\label{sec:data}

\subsection{Data Sources}

Our primary data source is the Hospital Episode Statistics Admitted Patient Care (HES APC) database, maintained by NHS England, which records all NHS-funded inpatient and day-case activity in England \citep{herbert2017data}. Each record contains patient-level information including diagnoses (ICD-10), procedures (OPCS-4), admission type, and hospital.\footnote{In the English NHS, OPCS-4 is a Fundamental Information Standard and the mandated clinical classification system used to codify operations, procedures, and interventions performed on patients. While diagnoses are recorded using ICD-10 codes, actual surgical procedures are recorded using OPCS-4 codes.} 

We focus on elective HVLC procedures across specialties targeted by the surgical hub programme, identified using OPCS-4 coding specifications published by the GIRFT programme. We retain elective inpatient and day-case admissions containing at least one qualifying HVLC procedure code. To account for procedure-level variations in resource intensity, each HVLC admission is linked to its national reference unit cost at the Healthcare Resource Group (HRG) level from the NHS National Cost Collection.\footnote{Healthcare Resource Groups (HRGs) are standard, clinically similar groupings of patient treatments that consume similar levels of NHS resources, analogous to Diagnosis Related Groups (DRGs). Derived from patient diagnoses and procedures, they enable fair reimbursement, cost benchmarking, and casemix management \hyperlink{https://digital.nhs.uk/services/national-casemix-office/the-why-what-and-how-of-casemix/the-casemix-companion/healthcare-resource-groups}{(NHS England, 2023)}.} 

Our secondary data sources are the Electronic Staff Record (ESR) and the NHS payroll system, which capture hospital staffing and salary. We extract monthly full-time equivalent (FTE) senior physician staffing levels and salary data at the hospital--specialty level.

To express financial measures in real terms, we draw official price deflator series from NHS England namely the NHS Cost Inflation Index for procedure unit costs, and the Health and Care Pay Cost Index (HCHS Pay Index) for physician salaries. Finally, information on hub adoption dates, hub configuration, and specialty coverage is obtained from an administrative register maintained by GIRFT, NHS England. We merge these disparate sources to create a consolidated panel dataset of elective outputs, physician inputs, and hub adoption status. Accordingly, the primary unit of analysis is the hospital–month, spanning the period from April 2014 to March 2024.

\subsection{Sample Construction}

The analysis sample comprises 40 NHS hospitals that adopted at least one HVLC surgical hub between April 2020 and March 2024 (the treated group) and 61 NHS hospitals that did not adopt a hub at any point during the study period (the never-treated control group), for a total of 101 hospitals and 11,705 hospital-month observations.

We exclude independent sector hospitals treating NHS-funded patients, as these operate under different contractual and organisational arrangements and are not subject to the hub programme. We exclude NHS hospitals with pre-existing hub-like arrangements established before April 2020, as these predate the formal programme and may reflect different organisational objectives and resource contexts. For treated hospitals with multiple hubs, treatment onset is defined by the earliest hub opening date, consistent with the argument that the relevant shock to elective-emergency separation occurs when the first dedicated facility becomes operational.

The 120-month panel length is determined by data availability: April 2014 is the earliest month for which HES APC data are consistently linked to ESR at the hospital-specialty level, and March 2024 is the latest available at the time of analysis. This gives early adopters those opening hubs in April 2020 up to 72 months of pre-treatment data and up to 48 months of post-treatment data, providing a long pre-treatment window for the parallel trends test and sufficient post-treatment horizon to assess whether productivity gains are sustained.

\section{Conceptual Framework}
\label{sec:framework}

To examine the potential productivity effects of surgical hubs, we outline a simple framework illustrating how emergency care distorts elective efficiency. We then construct empirical indices for cost-weighted output, salary-weighted input, and real physician productivity. Throughout this section, subscript $h$ indexes hospitals, $m$ indexes months, $s$ indexes surgical specialties, $j$ indexes specific HRG procedures, and $t$ indexes financial years.

\subsection{A Production Function Framework}

Hospital output in a given period can be represented as a function of labour, capital, and organisational arrangements that govern how inputs are allocated across competing uses. For elective surgery specifically, the relevant labour input is senior physician time—a scarce, expensive, and highly specialised factor. Let $X_{ht}^{E}$ denote elective output (measured in cost-weighted procedure volume) produced by hospital $h$ in period $t$, and let $Z_{ht}$ denote total senior physician FTE available. Physician productivity in elective care is:

\begin{equation}
Y_{ht}^{E} = \frac{X_{ht}^{E}}{Z_{ht}^{E}}
\end{equation}

In a mixed acute hospital, physician time ($Z_{ht}$) is divided between elective and emergency work, although observational data do not allow us to directly trace how individual physicians split their time between $Z_{ht}^{E}$ and $Z_{ht}^{A}$ on a day-to-day basis:

\begin{equation}
Z_{ht} = Z_{ht}^{E} + Z_{ht}^{A}
\end{equation}

where $Z_{ht}^{E}$ is physician time allocated to elective procedures and $Z_{ht}^{A}$ is time absorbed by emergency admissions. 

The economic problem is that $Z_{ht}^{A}$ is stochastic and responds to emergency demand shocks that are outside the hospital's control. When an emergency surge occurs, physicians are redeployed from elective lists—planned procedures are canceled or rescheduled, and and the physician time available for elective production falls below its contracted level. The difference between true and measured elective productivity is given by: 

\begin{equation}
\Delta Y_{ht}^{E} = \frac{X_{ht}^{E}}{Z_{ht}^{E}} - \frac{X_{ht}^{E}}{Z_{ht}} = X_{ht}^{E} \cdot \frac{Z_{ht}^{A}}{Z_{ht}^{E} \cdot Z_{ht}}
\end{equation}

which is increasing in $Z_{ht}^{A}$, the emergency burden, and decreasing in $Z_{ht}^{E}$, the elective capacity share.

The surgical hub programme can be understood as an investment in new organisational structures that reduces the stochastic reallocation of $Z^{E}$ toward emergency work. By ring-fencing theatres, wards, and physician rotas from the emergency pathway, a hub creates an institutional barrier between the two input pools. If the hub is effective at maintaining strict physical and operational separation, $Z_{ht}^{E}$ becomes protected from emergency demand shocks, reducing the influence of $Z_{ht}^{A}$ and causing measured elective productivity to rise. In the extreme case of complete separation—where physicians working in the hub are fully ring-fenced from emergency duties—the covariance between $Z_{ht}^{A}$ and $Z_{ht}^{E}$ approaches zero. This provides a theoretical basis for expecting larger productivity gains from standalone hubs relative to ring-fenced or integrated configurations.

\subsection{Productivity Measurement}

Our primary outcome is a measure of real physician productivity, defined as the ratio of real cost-weighted HVLC output to real salary-weighted physician input. Formally, we define our preferred productivity index, $Y_{hm}^{\text{CW-FTE-S}}$, as:

\begin{equation}
\label{eq:productivity definition}
  Y_{hm}^{\text{CW-FTE-S}} = \frac{
    \displaystyle X_{hm}^{CW} 
  }{
    \displaystyle Z_{hm}^{FTE-S} 
  }
\end{equation}

Where $X_{hm}^{CW}$ represents the cost-weighted elective  output and $Z_{hm}^{FTE-S}$ denotes the salary-weighted physician input. We measure the raw elective output simply as the total number of qualifying HVLC procedures performed by hospital $h$ in month $m$.

\subsubsection{Cost-Weighted Output Index}
We apply cost weights to outputs because surgical procedures differ fundamentally in their clinical complexity and resource intensity; using raw procedure counts would treat all HVLC activity as homogeneous. National average unit costs ($\bar{c}_{jst}$), deflated to 2019/20 base prices, are used rather than local costs so that output reflects real resource intensity rather than local pricing variations:

\begin{equation}
\label{eq:Pay and Price NHS CCI}
\bar{c}_{jst} = \frac{c_{jst}}{1 + \text{Infl}_{t}^{PP}}
\end{equation}

where $c_{jst}$ is the nominal national unit cost for procedure $j$ in specialty $s$, and $\text{Infl}_{t}^{PP}$ is the NHS Pay and Price Cost Inflation Index. Aggregating across procedures and specialties yields real cost-weighted output $X_{hm}^{CW}$:

\begin{equation}
\label{CW HVLC output}
 X_{hm}^{CW} = \displaystyle\sum_{s} \sum_{j} x_{jshm} \cdot \bar{c}_{jst}
\end{equation}

where $x_{jshm}$ is the volume of HVLC procedure $j$ in specialty $s$ delivered by hospital $h$ in month $m$.

\subsubsection{Salary-Weighted Input Index}
We apply salary weights to physician inputs because compensation reflects differences in the marginal productivity of labour across physician grades and specialties. Let $z_{shm}^{FTE}$ denote the total FTE senior physicians employed in specialty $s$, hospital $h$, and month $m$. Because physician FTE is recorded at the total specialty level, we apply an apportionment factor ($\phi_{shm}^{CW}$)—the share of cost-weighted HVLC output relative to total cost-weighted specialty output—to isolate the portion of physician time plausibly devoted to HVLC work.

Each apportioned FTE is weighted by the deflated national average annual salary ($\bar{\omega}_{st}$) for a senior physician in specialty $s$ during year $t$:

\begin{equation}
\label{eq:Pay NHS CCI}
\bar{\omega}_{st} = \frac{\omega_{st}}{1 + \text{Infl}_{t}^{Pay}}
\end{equation}

where $\text{Infl}_{t}^{Pay}$ is the Health and Care Pay Cost Index. Real salary-weighted physician input $Z_{hm}^{FTE-S}$ is defined as:

\begin{equation}
\label{Physician input weighted by salary}
   Z_{hm}^{FTE-S} = \displaystyle\sum_{s} \phi_{shm}^{CW} \cdot z_{shm}^{FTE} \cdot \bar{\omega}_{st}
\end{equation}

The resulting productivity metric $Y_{hm}^{\text{CW-FTE-S}}$ is interpreted as the real value of HVLC output produced per real pound of senior physician input cost. Expressing both numerators and denominators in constant 2019/20 prices ensures that measured productivity reflects changes in real volume and case-mix per unit of labour input rather than nominal price or wage inflation.

Beyond these main specifications, we examine sensitivity to how physician time is allocated to HVLC work varying both the apportionment rule and the input unit, and to the cost adjustment approach. Full details are provided in Appendix \ref{app:allocation} and the corresponding estimates are discussed in Section \ref{sec:robustness}.

\section{Empirical Strategy} \label{sec:empirics}

\subsection{Identification}

To estimate the average causal effect of surgical hub adoption on physician productivity we exploit the staggered rollout of hubs across hospitals and over time, which generates quasi-experimental variation in treatment status.  
Let $D_{hm} = 1$ if hospital $h$ has adopted a hub by month $m$, and let $E_h$ denote the month of first hub adoption for treated hospitals ($E_h = \infty$ for never-treated hospitals). The treatment effect for hospital $h$ in month $m$ is:

\begin{equation}
\tau_{hm} = Y_{hm}(1) - Y_{hm}(0)
\end{equation}

where $Y_{hm}(1)$ is the potential productivity outcome under hub adoption and $Y_{hm}(0)$ is the counterfactual outcome that would have been realised absent adoption. Since we observe only one potential outcome per hospital-month, identification requires imputing $Y_{hm}(0)$ for treated observations. This imputation rests on two assumptions. The first is \textit{parallel trends}: in the absence of hub adoption, treated hospitals would have followed the same productivity trajectory as never-treated hospitals. Formally,

\begin{equation}
\mathbb{E}[Y_{hm}(0) - Y_{hm'}(0) \mid E_h < \infty] 
= 
\mathbb{E}[Y_{hm}(0) - Y_{hm'}(0) \mid E_h = \infty]
\quad \forall\ m, m'.
\end{equation}

The second is \textit{no anticipation}: hospitals did not alter their elective scheduling or workforce practices in anticipation of hub opening, so pre-adoption productivity reflects business as usual rather than early behavioural responses to the forthcoming hub. These assumptions ensure that pre-adoption observations can be used to identify the counterfactual. 

We assess the plausibility of both assumptions empirically. The parallel trends assumption is assessed using a pre-trend test on untreated observations, described below. We test the no-anticipation assumption by re-estimating the model using placebo treatment dates assigned 6 and 9 months before actual hub opening and verifying that the placebo ATT is statistically indistinguishable from zero.

\subsection{Estimator}

Treatment effects are likely to vary across treated hospitals, reflecting differences in hub type, hospital size, or adoption timing. In such settings, the traditional two-way fixed effects (TWFE) estimator can assign negative weights to some treated observations, producing biased estimates in sign as well as magnitude \citep{goodman-bacon_difference--differences_2021}. We therefore adopt the imputation estimator of \citet{borusyak_revisiting_2024} (hereafter BJS) as our primary estimator, which is explicitly designed for staggered adoption with heterogeneous treatment effects.

BJS estimates the counterfactual  $\hat{Y}_{hm}(0)$ by fitting a two-way fixed effects model on untreated observations: never-treated units in all periods and treated units in their pre-adoption periods, and imputes the counterfactual for each treated hospital-month using the estimated hospital ($\hat{\alpha}_h$) and month fixed effects ($\hat{\beta}_m$) together with the estimated covariate coefficients 
applied to that hospital's own observed case-mix values (${W}'_{hm}$) in the post-adoption period:

\begin{equation}
\hat{Y}_{hm}(0) = \hat{\alpha}_h + \hat{\beta}_m + \hat{\varphi} {W}'_{hm}
\end{equation}

Where ${W}'_{hm}$  is a vector of time-varying patient case-mix controls including the proportion of patients with four or more comorbidities, average patient age, and the proportions of male, Asian, Black, and unknown-ethnicity patients.\footnote{Note that weighting elective output by HRG national average unit costs already standardizes baseline procedure-level resource intensity and clinical complexity. However, we include the vector $W'_{hm}$ to capture residual, within-HRG patient risk factors that remain distinct from procedure-level tariffs. Because national cost weights reflect broad, discrete average costs across all hospital inputs (including overheads and consumables), granular patient-level controls account for continuous variations in case severity and physician time intensity that fixed HRG tariffs maynot fully absorb. We include these controls to test their marginal impact on our estimates, and we also report parallel specifications without case-mix controls to confirm our findings are not driven by these adjustments.}
 
The estimated treatment effect for each treated hospital-month is then:

\begin{equation}
\hat{\tau}_{hm} = Y_{hm} - \hat{Y}_{hm}(0) = Y_{hm} - \hat{\alpha}_h - \hat{\beta}_m - \hat{\varphi}{W}'_{hm}
\end{equation}

and the average treatment effect on the treated (ATT) is the weighted average of $\hat{\tau}_{hm}$ across all treated hospital-month observations. 

Several heterogeneity-robust alternatives to TWFE exist, including \citet{callaway_difference--differences_2021} (CS). Among these, we prefer BJS for reasons specific to our data structure, namely that our pre-treatment panels are long and adoption cohorts are small. \citet{roth_whats_2023} show that BJS tends to be more efficient than CS when errors are not too serially correlated and parallel trends holds over a long pre-treatment periods, conditions our pre-trend test supports, as shown in the appendix. We discuss the data-specific reasons for preferring BJS over CS and verify this empirically in Section \ref{sec:robustness}, where we also benchmark against TWFE, which remains valid under homogeneous treatment effects. 

\subsection{Estimation}

\textbf{ATT specification.} The average treatment effect on the 
treated, $\tau^{ATT}$, is formally the parameter recovered when 
$Y_{hm}$ is expressed as:

\begin{equation} \label{eq:att}
Y_{hm} = \alpha_h + \beta_m + \tau^{ATT} D_{hm} + \varphi{W}'_{hm} + \varepsilon_{hm},
\end{equation}

where $\alpha_h$ are hospital fixed effects capturing time-invariant heterogeneity, $\beta_m$ are month fixed effects capturing common month-specific shocks including the COVID-19 shock to elective activity, $D_{hm}$ is the treatment indicator, and ${W}'_{hm}$  is a vector of time-varying patient case-mix controls defined above. 
We do not estimate equation~\eqref{eq:att} directly by OLS; rather, $\hat{\tau}^{ATT}$ is recovered as the precision-weighted average of the unit-level imputation residuals 
$\hat{\tau}_{hm}$ described above, following \citet{borusyak_revisiting_2024}. 
Equation~\eqref{eq:att} is estimated directly by OLS only in the TWFE robustness check reported in Section~\ref{sec:robustness}. Standard errors are clustered at the hospital level throughout.\footnote{We implement the BJS estimator using the \texttt{did\_imputation} Stata command \citep{borusyak_revisiting_2024}. By default, \texttt{did\_imputation} computes standard errors using cohort-by-month cells to form the comparison group. We assess robustness using the \texttt{leaveout} option with \texttt{avgeffectsby(D\_post)} for the overall ATT, as recommended by BJS, to ensure that no small cohort disproportionately influences the variance. Results are robust to this alternative variance specification.}

\textbf{Event study specification.} The dynamic treatment effects 
$\tau_k^{ES}$ are formally the parameters recovered when $Y_{hm}$ 
is expressed as:

\begin{equation} \label{eq:es}
Y_{hm} = \alpha_h + \beta_m + \sum_{k=0}^{K} \tau_k^{ES} D_{hm}^{k} + \varphi{W}'_{hm} + \varepsilon_{hm},
\end{equation}

where $D_{hm}^{k}$ equals one if hospital $h$ is $k$ months 
post-adoption in month $m$. As with $\tau^{ATT}$, we do not 
estimate equation~\eqref{eq:es} directly by OLS; $\hat{\tau}_k^{ES}$ is recovered by averaging the unit-level imputation residuals $\hat{\tau}_{hm}$ separately within each horizon $k$. We report estimates for 
$k = 0, \ldots, 12$, the horizon over which a sufficient number of treated hospitals contribute to each event-time estimate. At least 30 treated hospitals contribute to every estimate up to $k = 10$; the effective sample begins to decline at $k = 11$ and $k = 12$ as late adopters, whose post-treatment window is short relative to the study end date, exit the estimation window. Estimates beyond $k = 12$ are reported in the appendix for completeness.

\textbf{Pre-trend test.} Following BJS, we test whether productivity was trending in a similar fashion between eventually treated and those untreated using untreated observations separately:

\begin{equation} \label{eq:pretrend}
Y_{hm} = \alpha_h + \beta_m + \sum_{k \in \mathcal{K}_{\text{pre}}} \delta_k \text{Lead}_{hm}^{(k)} + \varphi{W}'_{hm} + \varepsilon_{hm},
\end{equation}

where $\text{Lead}_{hm}^{(k)}$ equals one if hospital $h$ is $k$ periods prior to adoption in month $m$. The set $\mathcal{K}_{\text{pre}}$ includes the 12 months prior to adoption in the baseline specification and all available  months in an extended check. Under parallel trends, $\delta_k = 0$ for all $k$. We test this using a joint $\chi^2$ test.

A potential threat to identification is that the COVID-19 pandemic disrupted hospital productivity differentially across hospitals, and hub adoption timing from April 2020 onward is contemporaneous with the pandemic. We address this concern through three complementary approaches. First, month fixed effects $\beta_m$ absorb any common time shock to productivity, including the aggregate suppression of elective volumes during lockdown periods. Second, we augment the main model with interactions between NHS England region dummies and COVID wave indicators corresponding to the first, Alpha, Delta, and Omicron waves to absorb region-specific wave shocks that may have differentially affected hospitals depending on their geographic location. Third, Section \ref{sec:robustness} reports results excluding the COVID disruption window (April 2020 to March 2021) entirely.


\section{Results}
\label{sec:results}

\subsection{Descriptive Statistics}

Table \ref{tab:desc} reports descriptive statistics for the productivity measure and key controls, disaggregated by treatment status and pre/post-adoption period. Pre-period averages for treated units are computed relative to each hospital's own adoption date, accounting for the variation in treatment timing across cohorts; control units use the same pre/post split as the treated unit to which they are most comparable in the pre-period.

Two features of Table \ref{tab:desc} are worth noting. First, pre-period means are closely comparable across treated and control groups in terms of productivity (6.30 vs 6.78 cost-weighted units of HVLC activity per physician). The post-period decline in productivity (3.68 vs 4.11) in both groups may reflect the COVID-19 shock to elective activity; this common shock is absorbed by month fixed effects in all specifications. Second, treated hospitals are substantially larger than control hospitals in terms of the FTEs of senior physicians employed (60.7 versus 41.7 FTEs), which may suggest that larger hospitals were more likely to adopt hubs, consistent with the scale requirements of the programme.  

\begin{table}[!htb]
\centering
\caption{Descriptive Statistics}
\label{tab:desc}
{\small
\begin{tabular}{lcccccc}
\toprule
 & \multicolumn{3}{c}{\textbf{Treated ($N = 40$)}} 
 & \multicolumn{3}{c}{\textbf{Control ($N = 61$)}} \\
\cmidrule(lr){2-4}\cmidrule(lr){5-7}
 & Mean & SD & Obs. & Mean & SD & Obs. \\
\midrule
\multicolumn{7}{l}{\textit{Primary outcome}} \\[2pt]
CW-HVLC / Wage-FTE \textit{(pre)}  
  & 6.30 & 3.42 & 3,676 
  & 6.78 & 2.47 & 4,183 \\
CW-HVLC / Wage-FTE \textit{(post)} 
  & 3.68 & 1.71 & 1,088 
  & 4.11 & 2.14 & 2,758 \\[6pt]
\multicolumn{7}{l}{\textit{Labour input}} \\[2pt]
Physician FTE \textit{(pre)}  
  & 60.67 & 33.22 & 3,676 
  & 41.67 & 18.74 & 4,183 \\
Physician FTE \textit{(post)} 
  & 67.87 & 37.16 & 1,088 
  & 38.79 & 22.48 & 2,758 \\[6pt]
\multicolumn{7}{l}{\textit{Patient characteristics (pre-period)}} \\[2pt]
\% of male patients       
  & 61.35 & 5.56 & 3,676 
  & 61.90 & 9.58 & 4,183 \\
Average patient age    
  & 60.72 & 5.58 & 3,676 
  & 61.84 & 7.34 & 4,183 \\
\% of multimorbid patients ($\geq$4 conditions)
  & 0.62 & 1.81 & 3,676 
  & 0.61 & 1.30 & 4,183 \\
\% of Asian patients     
  & 6.24 & 7.44 & 3,676 
  & 2.97 & 5.72 & 4,183 \\
\% of Black patients     
  & 3.89 & 5.39 & 3,676 
  & 1.34 & 3.88 & 4,183 \\
\% of unknown/not given ethnicity   
  & 16.42 & 12.93 & 3,676 
  & 14.13 & 13.25 & 4,183 \\[6pt]
\multicolumn{7}{l}{\textit{Patient characteristics (post-period)}} \\[2pt]
\% of male patients       
  & 63.22 & 6.59 & 1,088 
  & 62.38 & 12.17 & 2,758 \\
Average patient age    
  & 61.39 & 6.17 & 1,088 
  & 61.96 & 8.53 & 2,758 \\
\% of multimorbid patients ($\geq$4 conditions)
  & 0.83 & 0.77 & 1,088 
  & 0.81 & 2.21 & 2,758 \\
\% of Asian patients     
  & 7.55 & 7.36 & 1,088 
  & 3.31 & 6.20 & 2,758 \\
\% of Black patients     
  & 4.34 & 5.35 & 1,088 
  & 1.47 & 4.51 & 2,758 \\
\% of unknown/not given ethnicity   
  & 18.46 & 14.27 & 1,088 
  & 15.64 & 13.11 & 2,758 \\
\bottomrule
\multicolumn{7}{p{14cm}}{\footnotesize \textit{Notes:} 
Unit of observation is the hospital-month. $N = 40$ and $N = 61$ 
refer to the number of treated and never-treated hospitals 
respectively. Pre = before own adoption date for treated hospitals; 
before April 2020 for never-treated hospitals. Post = from adoption date 
onward for treated hospitals; from April 2020 onward for never-treated 
hospitals. CW = cost-weighted. Wage-FTE = salary-weighted physician FTE. 
Both output and input deflated to 2019/20 prices using the NHS cost 
inflation index and HCHS pay index, respectively.} \\
\end{tabular}
}
\end{table}

\clearpage

\begin{figure}[!ht]
   \centering
    \includegraphics[width=\linewidth]{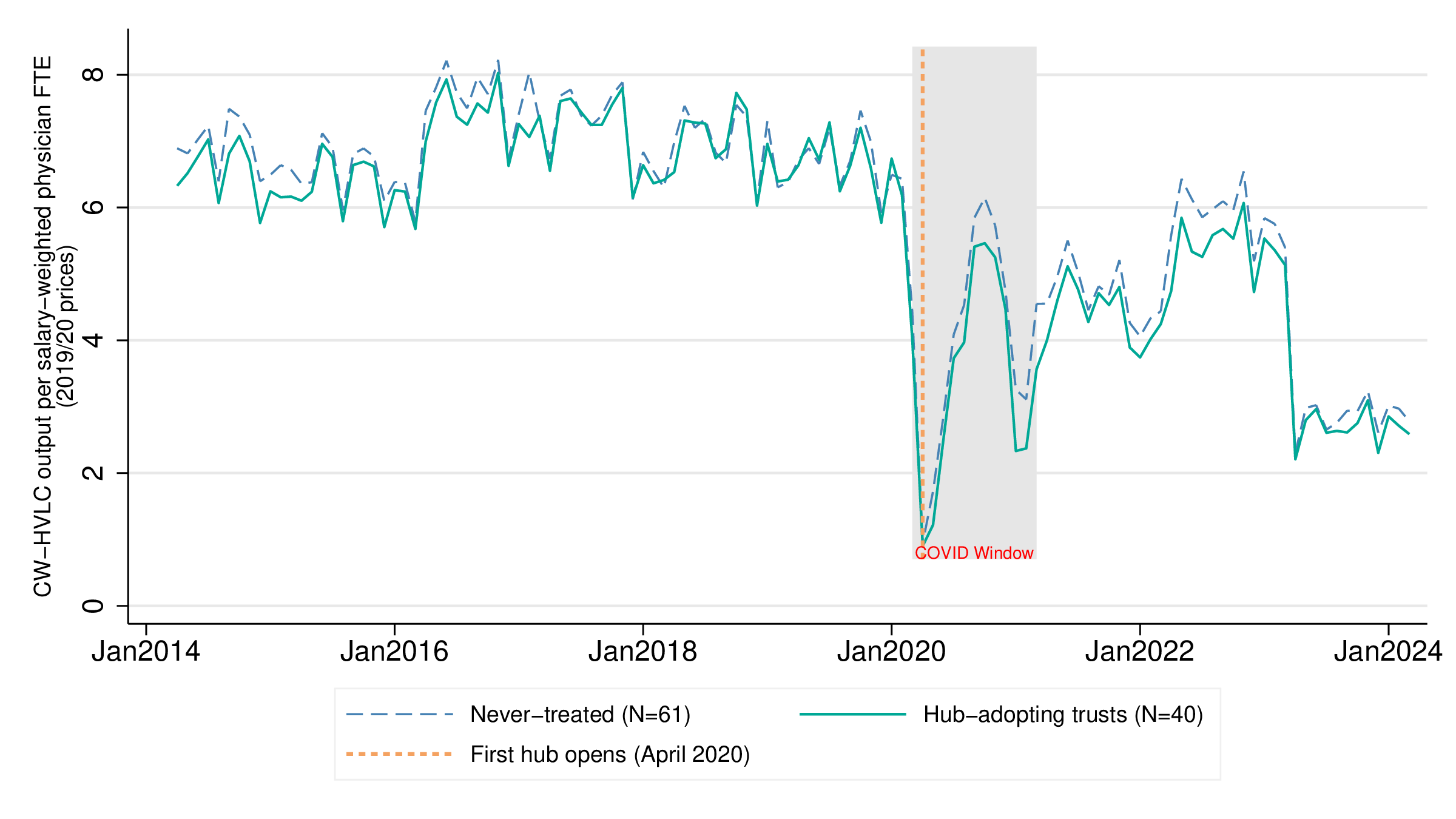}
    \caption{Trends in physician productivity}
    \label{fig:trends}
  \end{figure}

Figure \ref{fig:trends} plots monthly average cost-weighted HVLC output per salary-weighted physician FTE for hub-adopting and never-treated hospitals separately, from April 2014 to March 2024. Three features stand out. First, the two groups follow nearly identical productivity trajectories over the six years prior to the opening of the first hub (vertical dashed line), consistent with the parallel trends assumption that underpins our identification. Second, both groups experienced a sharp and simultaneous collapses in productivity during the COVID-19 pandemic. Third, although the two series remain broadly similar in levels, the post-adoption period show greater divergence between adopting and non-adopting hospitals relative to the pre-programme period. While these descriptive patterns are suggestive, the difference-in-differences estimates presented in Section \ref{sec:results} formally quantify these effects while accounting for the staggered timing of the adoption of hubs across hospitals.

\clearpage

\subsection{Main Results}
\label{sec:main_results}

Table \ref{tab:main} reports estimates of $\widehat{\tau}^{\mathrm{ATT}}$ across three productivity specifications, each estimated with and without patient case-mix covariates. Presenting these side-by-side allows us to assess whether the estimated hub effect is robust both to the choice of productivity measure and to the inclusion of case-mix adjustments. Columns (5) and (6) report our preferred specification--cost-weighted HVLC output per salary-weighted apportioned FTE, both deflated to 2019/20 prices. 

Across all six specifications, surgical hub adoption is associated with a statistically significant increase in physician productivity. Within each pair of outcome, the ATT is stable to the inclusion of case-mix covariates, and the effect remains significant as each cost-weighting or salary-weighting layer is added. Because the three outcomes are expressed in different units, magnitudes are not directly comparable across them; what matters is that sign and significance are preserved throughout.

The physician productivity measure in Column (6) can be interpreted as the monetary value of elective HVLC output per the monetary value of physician input. The ATT of 0.468 means that hub-adopting hospitals produced additional HVLC output worth £0.47 per £1 worth of real physician input, an increase of approximately 14.5\% relative to what they would have produced in the absence of hub adoption. This difference is identified by comparing each treated hospital to its own imputed counterfactual, holding fixed all time-invariant hospital characteristics and common time trends.

This percentage gain is recovered from the counterfactual mean, the productivity level treated hospitals would have achieved had they not introduced a surgical hub, as:

\begin{equation}
\bar{Y}^{\text{CF}} = \bar{Y}^{\text{post}}_{\text{treated}} - 
\widehat{\tau}^{\mathrm{ATT}} = 3.685 - 0.468 = 3.217
\end{equation}

where $\bar{Y}^{\text{post}}_{\text{treated}} = 3.685$ (rounded; 3.68 in Table~\ref{tab:desc}) is the observed average productivity of treated hospitals across all post-adoption months. Hub adoption therefore raised physician productivity from an estimated counterfactual of \pounds3.217 per \pounds1 worth of physician input to an observed \pounds3.68, a gain of approximately 14.5\% over the no-hub baseline. Although the BJS estimator does not directly report the counterfactual mean, we recover it from the definition of the ATT and it is identified by the same assumptions that identify the ATT itself. 

\begin{table}[H]
\centering
\caption{The Impact of Surgical Hub Adoption on Physician Productivity}
\label{tab:main}
{\small
\resizebox{\columnwidth}{!}{%
\begin{tabular}{lcccccc}
\toprule
 & (1) & (2) & (3) & (4) & (5) & (6) \\[2pt]
 & \multicolumn{2}{c}{HVLC / FTE} 
 & \multicolumn{2}{c}{CW-HVLC / FTE} 
 & \multicolumn{2}{c}{CW-HVLC / Wage-FTE} \\
\cmidrule(lr){2-3}\cmidrule(lr){4-5}\cmidrule(lr){6-7}
 & No controls & Controls 
 & No controls & Controls 
 & No controls & Controls \\
\midrule
$\widehat{\tau}^{\mathrm{ATT}}$ 
  & 0.495$^{***}$  & 0.507$^{***}$  
  & 3{,}463.7$^{***}$ & 3{,}583.2$^{***}$ 
  & 0.451$^{***}$ & 0.468$^{***}$ \\
  & (0.187) & (0.187) 
  & (1{,}193.4) & (1{,}200.5) 
  & (0.161) & (0.162) \\[6pt]
Counterfactual mean    & --- & --- & --- & --- & --- & 3.217 \\
\% above counterfactual & --- & --- & --- & --- & --- & 14.5\% \\[6pt]
Hospital FE   
  & \checkmark & \checkmark 
  & \checkmark & \checkmark 
  & \checkmark & \checkmark \\
Month FE      
  & \checkmark & \checkmark 
  & \checkmark & \checkmark 
  & \checkmark & \checkmark \\
Case-mix controls 
  &  & \checkmark 
  &  & \checkmark 
  &  & \checkmark \\
Observations  
  & 11{,}705 & 11{,}705 
  & 11{,}705 & 11{,}705 
  & 11{,}705 & 11{,}705 \\
\bottomrule
\multicolumn{7}{p{16cm}}{\footnotesize \textit{Notes:} 
 Case‑mix controls comprise the proportion of patients with four or more comorbidities, average patient age, and the proportions of male, Asian, Black, and unknown‑ethnicity patients, all constructed over the HVLC patient population. Columns (1)–(2) use unweighted HVLC procedure counts per physician FTE. Columns (3)–(4) use cost-weighted HVLC procedure output per physician FTE. Columns (5) and (6) deflate both cost‑weighted output and salary‑weighted FTE to 2019/20 real terms using the NHS Cost Inflation Index and HCHS pay index respectively. Standard errors clustered at the hospital level in parentheses. \(^{*}p<0.10\), \(^{**}p<0.05\), \(^{***}p<0.01\). BJS  estimator.} \\
\end{tabular}%
}}
\end{table}


\subsection{Event Study}
\label{sec:event_study}

\begin{figure}[!htbp]
\centering
\includegraphics[width=0.8\textwidth]{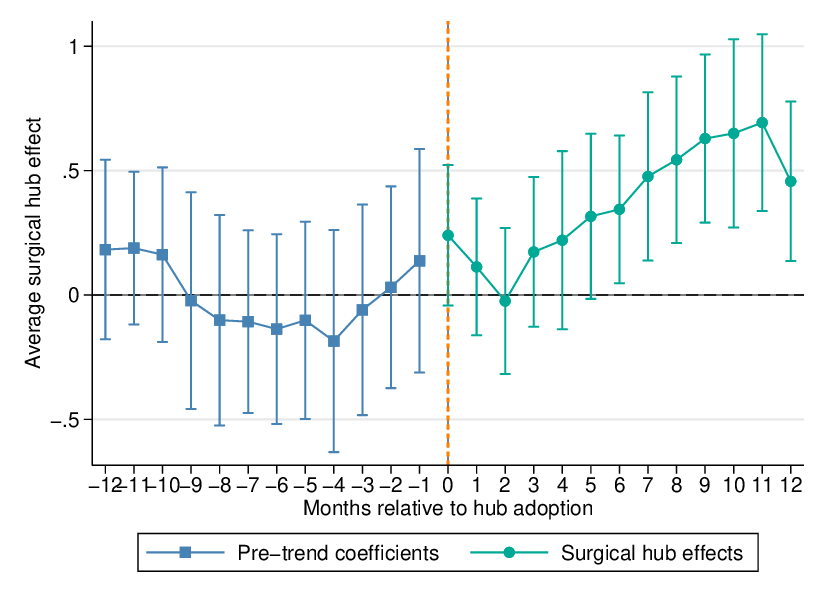}
\caption{\footnotesize Event study: dynamic effects of surgical hub adoption on physician productivity (CW‑HVLC / Wage‑FTE). Points are \(\widehat{\tau}_k^{\mathrm{ES}}\) for post‑adoption months (\(k\ge 0\), estimated on treated observations) and \(\widehat{\delta}_k\) for pre‑adoption months (\(k<0\), estimated on untreated observations). Error bars represent 95\% confidence intervals. Standard errors clustered at the hospital level. All specifications include hospital fixed effects, month fixed effects, and the full case‑mix control vector.}
\label{fig:es}
\end{figure}

Figure \ref{fig:es} presents the event study estimates for the post treatment periods based on equation \eqref{eq:es}, plotting \(\widehat{\tau}_k^{\mathrm{ES}}\) up to 12 months alongside pre‑trend coefficients \(\widehat{\delta}_k\) from equation \eqref{eq:pretrend}. 

The pre‑trend coefficients are statistically indistinguishable from zero across all 12 pre‑adoption months. A joint \(\chi^2\) test of the null that all 12 pre‑trend coefficients are jointly zero yields \(\chi^2(12) = 12\), \(p = 0.446\), failing to reject parallel trends.  We report pre-trend tests across all available pre-treatment windows in Appendix Figure \ref{fig:es_long1}; individual lead coefficients are statistically insignificant throughout, with substantially overlapping confidence intervals across them all.\footnote{However, the joint $\chi^2$ test over the full extended window rejects the null of no pre-trends at the 1\% level; Appendix~\ref{app:robustness} discusses why we believe this reflects over-rejection of long-window joint tests in finite samples and COVID-period noise rather than a substantive violation. Because such arguments cannot be decisive, the Supplementary Material (Section~\ref{S-sec:S2}) additionally reports Rambachan--Roth sensitivity bounds \citep{rambachan2023more}, which characterise how large a violation of parallel trends, relative to the largest pre-period violation between consecutive periods, would be required to overturn the statistical significance of the estimated ATT} 

Post-adoption event-study estimates suggest that productivity gains do not appear fully at hub opening, but emerge over the following months as the new organisational arrangements become established. From around three months after adoption, the estimates are consistently positive and point to a sustained improvement in physician productivity. 

Following the recommendation of \citet{borusyak_revisiting_2024}, we retain horizons where at least 30 treated units contribute to the estimate. This condition is met for all post-adoption horizons up to $k = 10$ and declines at $k = 11$ and $k = 12$. This is mainly due to late adopters whose post‑treatment observation window is short relative to the study end date. Estimates in the \(k=11\) to \(k=12\) range should therefore be interpreted with caution, though they remain based on a majority of the treated sample. 

Event study estimates for the full available post‑treatment horizons are reported in Appendix Figure \ref{fig:es_long0}. Most of these longer‑horizon estimates are individually significant and directionally consistent with the main window, suggesting that the productivity gains documented are stable. We give them, however, less interpretive weight given the declining sample size at extended horizons.

The overall pattern is difficult to reconcile with alternative explanations based on administrative reclassification or data recording changes at hub opening. A pure reclassification effect would produce a spike at \(k=0\) followed by immediate reversion to baseline, not the sustained and strengthening trajectory we observe from \(k=3\) onward. The dip at \(k=1\) and \(k=2\) is more naturally interpreted as the operational cost of transition the period during which the hub is open but scheduling, staffing, and patient flow have not yet fully adapted — after which the structural insulation of elective lists from emergency disruption generates the gains visible from \(k=3\) to \(k=12\) and beyond.

\clearpage
\subsection{No-Anticipation Test}
The no-anticipation assumption requires that hub adoption did not alter physician productivity before the hub became operational for instance through preparatory changes in scheduling, staffing, or patient selection in anticipation of the hub opening. We test this by assigning two separate placebo treatment dates, 6 and 9 months before the actual hub opening date, and re-estimating the ATT on the actual pre-treatment sample, restricting the estimation to observations prior to each hospital's actual hub opening.\footnote{We assign placebo treatment dates 6 and 9 months before actual hub opening. Shorter lags provide a focused test of anticipatory behaviour in the period immediately preceding adoption, where such responses are most plausible. Results are qualitatively unchanged if a 12-month lag is used instead.} If the no-anticipation assumption holds, the placebo ATT should be statistically indistinguishable from zero.

\begin{figure}[htbp]
\centering
\begin{subfigure}{0.48\textwidth}
\centering
\includegraphics[width=\textwidth]{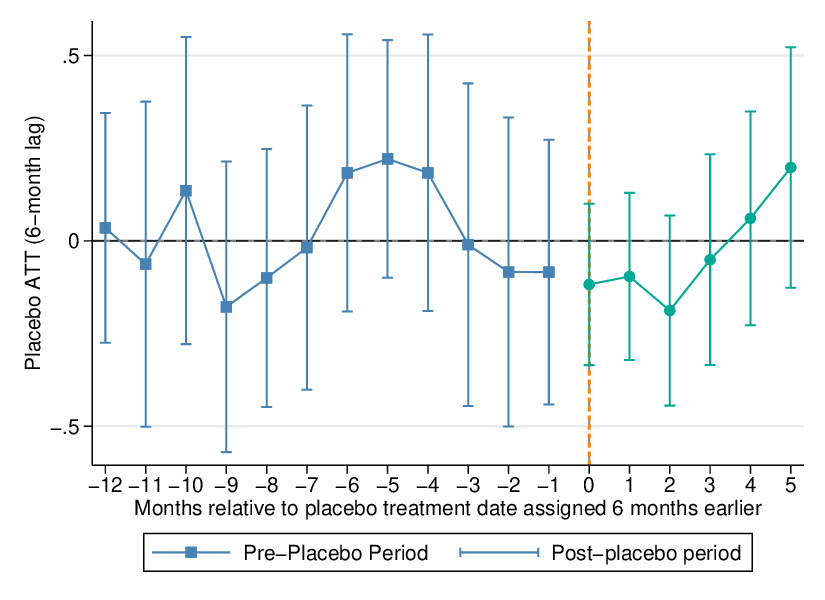}
\caption{Placebo date: 6 months before actual hub opening}
\label{fig:placebo6}
\end{subfigure}
\hfill
\begin{subfigure}{0.48\textwidth}
\centering
\includegraphics[width=\textwidth]{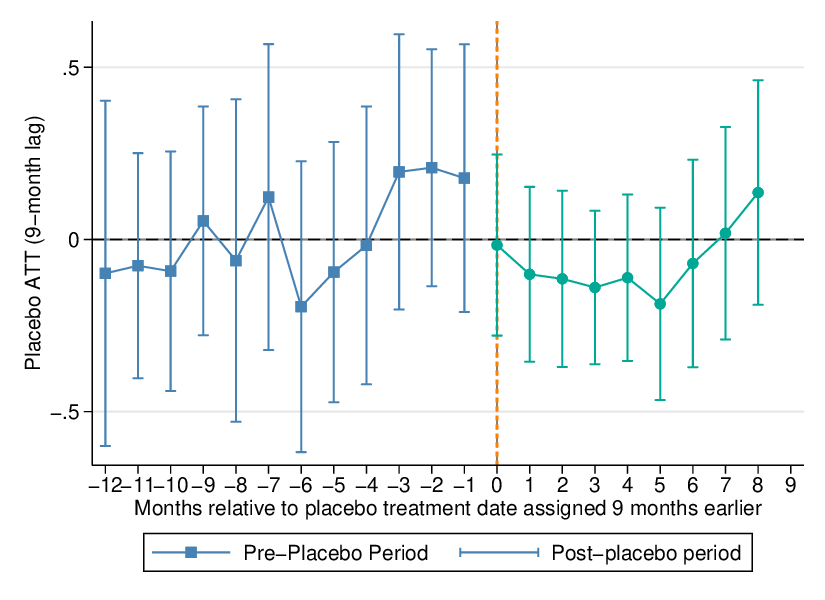}
\caption{Placebo date: 9 months before actual hub opening}
\label{fig:placebo9}
\end{subfigure}
\caption{No-anticipation test: placebo event study estimates. Points are monthly placebo ATT coefficients (\(\widehat{\tau}_k^{\text{placebo}}\)) for post-placebo months and pre-trend lead coefficients (\(\widehat{\delta}_k\)) for pre-placebo months. Estimation is restricted to observations prior to each hospital's actual hub opening date. Error bars represent 95\% confidence intervals with standard errors clustered at the hospital level. All specifications include hospital fixed effects, month fixed effects, and the full case-mix covariate vector. }
\label{fig:combined}
\end{figure}

Figure \ref{fig:combined} presents the placebo event study plots for both lags. In both cases, post-placebo coefficients are statistically indistinguishable from zero throughout, with confidence intervals spanning zero at every horizon. The aggregate placebo ATT is -0.049 (\(p=0.706\)) for the 6-month lag and \(-0.065\) (\(p=0.594\)) for the 9-month lag. 
A joint \(\chi^2(12)\) test fails to reject the null of no pre-placebo trend for both lags (\(\chi^2(12)=12.75\), \(p=0.387\) and \(\chi^2(12)=15.7\), \(p=0.207\), respectively). These results support the no-anticipation assumption.

\subsection{Heterogeneity analyses}
\label{sec:heterogeneity}

Table~\ref{tab:het} presents heterogeneity in the ATT across three dimensions of hub design: hub type (the depth of separation from the emergency pathway), hub specialisation (the number of specialties served), and the extent of adoption (the number of hubs operated). We call this last dimension extent of adoption rather than scale, because our data record how many separate hub facilities a hospital operates, not the volume of activity passing through them; measuring adoption intensity in that sense would require theatre utilisation data which, as noted in Section~\ref{sec:discussion}, we do not have. The Supplementary Material (Section~\ref{S-sec:S3}) reports the cross-tabulation of treated hospitals across categories, which shows the extent of overlap across the three dimensions and gives the number of hospitals contributing to each subgroup estimate in Table~\ref{tab:het}. Section~\ref{S-sec:S3} of the Supplementary Material also reports tests of equality of subgroup ATTs within each panel; given the small number of clusters within subgroups, these use a wild-cluster bootstrap alongside a conventional clustered Wald test.

\subsubsection{Hub type (Extent of separation)}

All three hub configurations deliver statistically significant gains, ranging from ($0.411$) for ring-fenced hubs to $0.532$ for standalone hubs, with integrated hubs in between ($0.450$). The ordering is consistent with gains increasing in the completeness of physical separation from the emergency pathway, though as the equality tests in the Supplementary Material (Section~\ref{S-sec:S3}) show, the three estimates are not statistically distinguishable from one another.

\subsubsection{Hub specialisation}

Hospitals operating single-specialty hubs show larger gains ($0.786$) than those operating multi-specialty hubs ($0.369$), and both are significant on their own. Whether this difference is itself significant depends on the estimator used, as the Supplementary Material (Section~\ref{S-sec:S3}) reports, and we therefore treat the specialisation contrast as unresolved rather than as evidence that specialisation amplifies the gain.

\subsubsection{Extent of adoption (number of hubs)}

Hospitals operating two or more hubs achieve gains of $1.110$, more than double the pooled average main effect ($0.468$), while single-hub adoption yields an effect of $0.239$ that is statistically indistinguishable from zero. Using a wild-cluster bootstrap to account for the small number of clusters in the two-or-more-hubs group, the Supplementary Material (Section~\ref{S-sec:S3}) confirms the difference is significant, consistently across estimators.

\begin{table}[H]
\centering
\caption{Heterogeneous Effects of Surgical Hub Adoption on 
Physician Productivity}
\label{tab:het}
{\small
\begin{tabular}{lccc}
\toprule
 & (1) & (2) & (3) \\
 & \multicolumn{3}{c}{CW-HVLC / Wage-FTE} \\
\cmidrule(lr){2-4}
 & Hub type & Specialisation & Scale \\
\midrule
\multicolumn{4}{l}{\textbf{Panel A: Hub type}} \\[2pt]
Integrated hub  
  & 0.450$^{**}$ & & \\
  & (0.192) & & \\[2pt]
Ring-fenced hub 
  & 0.411$^{**}$ & & \\
  & (0.207) & & \\[2pt]
Standalone hub  
  & 0.532$^{***}$ & & \\
  & (0.204) & & \\[4pt]
\multicolumn{4}{l}{\textbf{Panel B: Number of specialties served}} \\[2pt]
Multi-specialty hub 
  & & 0.369$^{**}$ & \\
  & & (0.163) & \\[2pt]
Single-specialty hub 
  & & 0.786$^{***}$ & \\
  & & (0.190) & \\[4pt]
\multicolumn{4}{l}{\textbf{Panel C: Number of hubs operated}} \\[2pt]
One hub        
  & & & 0.239 \\
  & & & (0.165) \\[2pt]
Two or more hubs 
  & & & 1.109$^{***}$ \\
  & & & (0.139) \\[4pt]
\midrule
Case-mix covariates 
  & \checkmark & \checkmark & \checkmark \\
Hospital FE     
  & \checkmark & \checkmark & \checkmark \\
Month FE        
  & \checkmark & \checkmark & \checkmark \\
Observations    
  & 11{,}705 & 11{,}705 & 11{,}705 \\
\bottomrule
\multicolumn{4}{p{10cm}}{\footnotesize
\textit{Notes:} Each column reports subgroup-specific ATT estimates 
obtained by aggregating unit-level imputation residuals separately 
for hospitals within each hub characteristic category. 
The counterfactual is estimated from a single pooled regression on 
untreated observations common to all specifications. 
The outcome variable is cost-weighted HVLC output per salary-weighted physician FTE, deflated to 2019/20 prices. Standard errors clustered at the hospital level in 
parentheses. $^{*}p<0.10$, $^{**}p<0.05$, $^{***}p<0.01$. BJS estimator.} \\
\end{tabular}
}
\end{table}

\subsubsection{Specialisation}

\label{sec:specialty}

This section examines whether the pooled productivity gain varies across the specialties targeted by the hub programme. Table \ref{tab:specialty} reports $\widehat{\tau}^{\mathrm{ATT}}$ separately for each specialty, using the preferred productivity measure defined in equation~\eqref{eq:productivity definition}. Hub adoption led to a statistically significant increase in physician productivity in T\&O ($0.604$) and General Surgery ($0.337$). 

The effects in the remaining specialties are positive, though statistically insignificant. That most specialties return positive point estimates is broadly consistent with the pooled finding. 
The stronger signal in T\&O is consistent with its dominant role in the treated sample: single-specialty hubs predominantly serve T\&O, and multi-specialty hubs almost always include T\&O procedures, meaning a larger share of treated hospitals contribute to the T\&O estimate.

In contrast, estimates for the remaining specialties reflect a combination of reduced statistical power and genuine null effects. For Ophthalmology, the point estimate is nearly as large as in T\&O  and clinically meaningful, but a smaller sample size yields a wider standard error ($0.344$) -- a power constraint especially evident in Urology, where the standard error rises to $0.831$. For Gynaecology $(0.226$) and ENT ($0.014$), point estimates are smaller, with ENT in particular returning an estimate near zero that suggests a true absence of productivity gains rather than a loss of precision.

\begin{table}[!htb]
\centering
\caption{Effects of Surgical Hub Adoption on Physician Productivity 
by Specialty}
\label{tab:specialty}
{\small
\resizebox{\columnwidth}{!}{%
\begin{tabular}{lcccccc}
\toprule
 & (1) & (2) & (3) & (4) & (5) & (6) \\
 & T\&O & Gen.\ Surgery & Gynaecology & Ophthalmology 
 & Urology & ENT \\
\midrule
$\widehat{\tau}^{\mathrm{ATT}}$ 
  & 0.604$^{**}$ & 0.337$^{**}$ & 0.226 
  & 0.561 & 0.270 & 0.014 \\
  & (0.271) & (0.160) & (0.217) 
  & (0.344) & (0.831) & (0.181) \\[6pt] \midrule
Case-mix covariates    
  & \checkmark & \checkmark & \checkmark 
  & \checkmark & \checkmark & \checkmark \\
Hospital FE   
  & \checkmark & \checkmark & \checkmark 
  & \checkmark & \checkmark & \checkmark \\
Month FE      
  & \checkmark & \checkmark & \checkmark 
  & \checkmark & \checkmark & \checkmark \\
Observations  
  & 11{,}011 & 11{,}032 & 11{,}301 
  & 9{,}377 & 10{,}775 & 9{,}505 \\
\bottomrule
\multicolumn{7}{p{15cm}}{\footnotesize \textit{Notes:} 
Each column reports a separate regression for the indicated 
specialty. The outcome variable is cost-weighted HVLC output per salary-weighted physician FTE, deflated to 2019/20 prices. Spinal surgery is excluded from this table because we cannot reliably estimate the productivity effect as the number of treated hospitals offering spinal HVLC procedures is small. Standard errors clustered at the hospital level in parentheses. $^{*}p<0.10$, $^{**}p<0.05$, $^{***}p<0.01$. 
BJS  estimator.} \\
\end{tabular}%
}}
\end{table}

\subsection{Robustness}
\label{sec:robustness}

Appendix~\ref{app:robustness} presents robustness checks. The baseline estimated effect (0.468) is stable across alternative DiD estimators with BJS and TWFE producing comparable statistically significant estimates, ($0.468$ and $0.425$ over the full window; $0.349$ and $0.341$ over the $k=0$--$12$ window), while Callaway--Sant\textquotesingle{}Anna estimate is positive but less precise over the full window ($0.183$, s.e.\ $0.175$), converging to the other estimators ($0.347$) in the restricted window.  Appendix~\ref{app:robustness} discusses the sources of this divergence. The baseline ATT is also little changed when we allow for region-specific COVID-wave shocks (0.446), exclude the acute COVID disruption window (0.455), omit the first one or two months after hub opening ($0.476; 0.491$), or vary the input allocation rule ($0.198-0.634$) and cost adjustment approach ($0.426-0.527$).

The Supplementary Material (Section~\ref{S-sec:S2}) reports further identification analyses motivated by the setting: Rambachan--Roth honest DiD sensitivity bounds on parallel trends \citep{rambachan2023more}, alongside a synthetic difference-in-differences estimate that relaxes the parallel-trends requirement itself rather than bounding the consequences of its violation \citep{arkhangelsky2021synthetic}; a test of whether pre-programme hospital characteristics predict adoption and its timing; robustness to the 2023--24 industrial action; and a check of whether the estimate depends on any single hospital or adoption cohort, together with trimming exercises addressing the sensitivity of the ratio outcome to extreme observations.

\section{Mechanisms}
\label{sec:mechanisms}

This section examines the channels through which hub adoption raises physician productivity. First, we examine the HVLC activity share which captures the fraction of total cost-weighted specialty activity accounted for by elective HVLC procedures, providing a direct measure of compositional reallocation toward the procedures the hub is designed to protect. Second, average waiting time and average length of stay which are patient-level outcomes aggregated to the hospital-month, capturing whether hub adoption affects access to and duration of the care episode. Third, the number of operating theatres and the number dedicated to day-case procedures which are obtained from quarterly NHS England administrative data with a caveat that this is only measured at the hospital level, not at specialty, and that data is missing for the COVID-19 pandemic periods due collection suspension. We use these last two outcomes as a proxy for physical capacity changes that could alternatively explain the productivity gain. Table \ref{tab:mechanism} reports the effects of hub adoption on these outcomes.

\textbf{Compositional shift toward HVLC activity.} Column (1) shows that hub adoption is associated with a statistically significant increase in the HVLC activity share ($0.009$), consistent with the conceptual framework: by insulating elective scheduling from emergency disruption, the hub raises the share of specialty activity devoted to HVLC work. To quantify how much of the overall productivity gain then operates through this compositional channel, we augments the baseline productivity specification with the HVLC activity share as an additional regressor (Table \ref{tab:mediation}).  The hub ATT falls from baseline $0.468$ to $0.390$, a difference of $0.078$, or approximately 17\% ($0.078/0.468$) of the baseline effect operates through the compositional shift toward HVLC activity, with the remainder reflecting other channels.\footnote{Note the HVLC activity share is itself an outcome of hub adoption, so conditioning on it raises a bad-controls problem \citep{angrist2009mostly}. The attenuation should therefore be interpreted as suggestive of partial mediation rather than a precise causal decomposition.}

\textbf{Patient waiting time and length of stay.} Column (2) shows that hub adoption reduces average patient waiting times by approximately 10 days (7.6\% lower than expected under the no-hub counterfactual). Column (3) shows no significant change in average length of stay following hub adoption. This is somewhat counter to what might be expected since hubs are designed to concentrate day-case activity, which would typically reduce average length of stay. 
\citet{wen2026effect} find some evidence of length of stay reduction for hip and knee replacement following hub adoption, though their estimates are subject to pre-trend concerns.

\textbf{Theatre capacity.} Columns (4) and (5) report the effects of the hub on the total number of operating theatres and the number of dedicated day-case theatres, respectively. Neither shows a statistically significant change. We note that hubs themselves may involve capital investment in dedicated facilities, and that theatre counts measured at the hospital level may not fully capture specialty-level changes in theatre availability. 

We additionally examine last-minute elective cancellations and the number of patients untreated within 28 days following cancellation. Again these data are collected on a quarter basis and the caveats mentioned above applies to this data. Particular to these outcomes, pre-trend coefficients show significant departures from zero across the pre-adoption window, precluding causal inference for these outcomes. We therefore report event study plots in Appendix \ref{app:cancellations} for completeness.

\begin{table}[!ht]
\centering
\caption{Mechanisms: Hub Effects on Intermediate Outcomes}
\label{tab:mechanism}
\resizebox{\columnwidth}{!}{%
\begin{tabular}{lccccc}
\hline
 & (1) & (2) & (3) & (4) & (5) \\
 & \begin{tabular}[c]{@{}c@{}}HVLC\\activity share\end{tabular}
 & \begin{tabular}[c]{@{}c@{}}Average\\waiting days\end{tabular}
 & \begin{tabular}[c]{@{}c@{}}Average\\length of stay\end{tabular}
 & \begin{tabular}[c]{@{}c@{}}Number of \\ operating theatres\end{tabular}
 & \begin{tabular}[c]{@{}c@{}}Dedicated day-case\\theatres\end{tabular} \\
\hline \\
$\widehat{\tau}^{\mathrm{ATT}}$
  & 0.009$^{**}$ & $-$10.316$^{**}$ & $-$0.009
  & 0.859 & 0.051 \\
  & (0.004) & (4.792) & (0.032)
  & (1.023) & (0.740) \\[6pt]
\hline
Case-mix covariates
  & \checkmark & \checkmark & \checkmark
  & \checkmark & \checkmark \\
Hospital FE
  & \checkmark & \checkmark & \checkmark
  & \checkmark & \checkmark \\
Month FE
  & \checkmark & \checkmark & \checkmark
  & \checkmark & \checkmark \\
Observations
  & 11{,}705 & 11{,}524 & 11{,}694
  & 3{,}247 & 3{,}247 \\
\hline
\multicolumn{6}{p{18cm}}{\footnotesize \textit{Notes:}
Each column reports the BJS ATT for a separate outcome variable.
Columns (1)--(3) use monthly hospital-level data. Columns (4) 
and (5) use quarterly hospital-level data from NHS England 
administrative records; the smaller sample reflects the 
availability of theatre data. Standard errors clustered at 
the hospital level in parentheses.
$^{*}p<0.10$, $^{**}p<0.05$, $^{***}p<0.01$. } \\
\end{tabular}%
}
\end{table}

\begin{table}[H]
\centering
\caption{Mediation Analysis: HVLC Activity Share}
\label{tab:mediation}
\resizebox{0.45\textwidth}{!}{%
\begin{tabular}{lc}
\hline
 & (1) \\
 & CW-HVLC / Wage-FTE \\
\hline \\
$\widehat{\tau}^{\mathrm{ATT}}$
  & 0.390$^{**}$ \\
  & (0.154) \\[4pt]
HVLC activity share
  & 8.99$^{***}$ \\
  & (1.42) \\[6pt]
\hline
Case-mix covariates & \checkmark \\
Hospital FE & \checkmark \\
Month FE & \checkmark \\
Observations & 11{,}705 \\
\hline
\multicolumn{2}{p{8cm}}{\footnotesize \textit{Notes:}
The outcome variable is cost-weighted HVLC output per salary-weighted physician FTE, deflated to 2019/2 prices. HVLC activity share is included as an additional regressor alongside the hub treatment indicator. Standard errors clustered at the hospital level in parentheses. $^{*}p<0.10$, $^{**}p<0.05$, $^{***}p<0.01$. BJS estimator.} \\
\end{tabular}%
}
\end{table}

\clearpage
\section{Discussion}
\label{sec:discussion}

A long-standing structural feature of hospital care is that scheduled elective work and unpredictable emergency demand compete for the same physicians, theatres, and beds. Elective activity is routinely displaced when emergency demand surges \citep{johar2013emergency, addison2001separating}. This paper asked whether separation of the two streams of activity, through dedicated elective hubs, allowed hospitals to use their physicians more productively. We found that it did; hub adoption raised physician productivity by approximately 14.5\% relative to the no-hub counterfactual, and the effect emerged gradually and was sustained.

The mechanism evidence points to the protection and concentration of elective activity as a channel for this gain. Following hub adoption, the share of specialty activity accounted for by HVLC elective work undertaken in the hospital rose, a compositional shift consistent with recent findings that hub adoption drives overall increases in elective surgical volume \citep{co_impact_2025, wen2026effect}. Conditioning on this compositional shift accounts for roughly 17\% of the total productivity effect, a decomposition we interpret as descriptive given that the activity share is itself an outcome of adoption. A higher HVLC activity share means that more elective output was delivered per unit of physician time, or more patients received treatment for a given level of staffing. Consistent with the rise in HVLC activity share, average waiting times fell by approximately ten days. One plausible explanation for the reduction in waiting times is that separation allowed scheduled lists to proceed with less interruption from emergency demand, so that protected physician time was converted into elective output more reliably. This pattern echoes earlier evidence that dedicating teams and facilities to emergency care allows elective lists to proceed uninterrupted and raises day-case throughput \citep{addison2001separating, parasyn2009acute, lowthian2011streamlining}. Average length of stay was unchanged. This is consistent with the nature of the HVLC procedures in our sample, which were already largely performed as day cases even before hub adoption. The organisational separation introduced by hubs does not materially alter the time these procedures inherently take. Rather, hubs appear to have increased throughput by enabling more HVLC cases to be completed; it seems not by shortening the already brief time these patients spent in hospital. Operating theatre counts were unchanged in our study, but this should be interpreted cautiously given the data limitations discussed earlier. Hub adoption was also financed through targeted programme capital and in some cases involved new or repurposed physical estate, so we do not claim that the productivity gain arises purely from the reorganisation of existing resources.

By estimating the impact of elective-emergency separation on physician productivity directly, our results bridge two distinct literatures. First, while recent evaluations document overall increases in surgical volumes following hub adoption \citep{co_impact_2025, wen2026effect}, we show that cost-weighted surgical output expanded per unit of physician input, consistent with genuine productivity gains. Second, our findings provide quantitative support for the focused factory hypothesis \citep{skinner1974focused, casalino2003focused, addison2001separating}. Prior studies show that specialized facilities like ambulatory surgery centers and ring-fenced units achieve lower procedure costs and shorter lengths of stay \citep{siciliani2013differences, barlow_effect_2013, munnich_returns_2018, hollenbeck_ambulatory_2015, carey_specialization_2019}. Our results demonstrate that this efficiency gain manifests directly as higher output per unit of physician time. Furthermore, that relatively the largest gains occur in standalone and single-specialty hubs reinforces the core prediction of this literature: the returns to specialization increase with the completeness of operational separation.

Some limitations should be borne in mind when interpreting our results. First, our productivity measure captures senior physician labour only. Resident physicians also contribute materially to hospital output \citep{perez2022contribution}. We lacked the data to observe how hub adoption affects the productivity of other staff or the division of labour between senior and junior staff. Second, the productivity measure we constructed reflects the real value of elective output per unit of senior physician input and is silent on most dimensions of clinical quality. Whether the gain was achieved without cost to clinical outcomes cannot be settled with the measures used here, and readmission, complication, and patient-reported outcomes remain an important direction for further work. Third, the hub registers, which record the operational status of hubs across NHS hospitals, record operational status but not the intensity with which hub theatres were used. Data on theatre utilisation would have allowed us to measure whether hubs were operating close to their capacity and whether any further productivity gains could be yielded by moving hubs closer to their maximum capacity.

\section{Conclusion}
\label{sec:conclusion}

A major health policy concern across many health systems is how to increase the productivity of hospital physicians while delivering scheduled elective treatment alongside unpredictable emergency demand within the same organisation. This paper examined one response to that tension: the organisational separation of elective and emergency surgical care, as implemented through NHS England's surgical hub programme.

A central insight emerging from our analysis is that organisational separation improves physician productivity across most hub configurations, though the magnitude of the gain depends on how complete and extensive the separation is. Gains are larger for standalone and single-specialty hubs, and largest for hospitals operating two or more hubs. Adoption of a single-hub was the one case that produced no discernible effect. The returns to separation therefore appear to depend on its depth and and the extent of adoption, rather than its mere presence.

Overall, the paper offers guidance to policy makers assessing how to organise elective care. For health systems confronting elective backlogs under fiscal constraints, the depth and scale of the organisational separation, rather than the simple fact of simply separating elective and emergency activity, emerges as the key determinant of whether the reorganisation improves physician productivity.

\section*{Disclaimer Statements}

\noindent \textbf{Acknowledgments:} We thank participants and our discussant, Silvana Robone, at the EuHEA seminar, and Karen Bloor for helpful discussions on earlier versions of this paper. The paper was presented at the HESG Brighton Conference and the EuHEA Rotterdam Conference.

\vspace{0.3cm}

\noindent \textbf{Funding:} This study is funded by the National Institute for Health Research, Health Services and Delivery Research Programme (NIHR HS\&DR), grant number 153387. The views expressed are those of the authors and not necessarily those of the NIHR or the Department of Health and Social Care.

\vspace{0.3cm}

\noindent \textbf{Data Statement:} This work uses data provided by patients and collected by the NHS as part of their care and support. Hospital Episode Statistics are copyright \textcopyright{} 2014/15--2023/24, NHS England. Electronic Staff Record data are provided by NHS England. Both are re-used with the permission of NHS England. All rights reserved. The data used in this study are not publicly available due to information governance restrictions; researchers seeking access should apply directly to NHS England.

\vspace{0.3cm}

\noindent \textbf{Conflicts of Interest:} None declared.

\bibliographystyle{agsm}
\bibliography{references}

\clearpage 

\section*{Appendix}
\appendix
\counterwithin{table}{section}
\counterwithin{figure}{section}
\renewcommand{\thetable}{\thesection\arabic{table}}
\renewcommand{\thefigure}{\thesection\arabic{figure}}
\renewcommand{\theequation}{\thesection\arabic{equation}}

\begin{appendices}

\section{Surgical Hub Effect: All post treatment periods} 

\begin{figure}[!ht]
   \centering
    \includegraphics[width=0.8\linewidth]{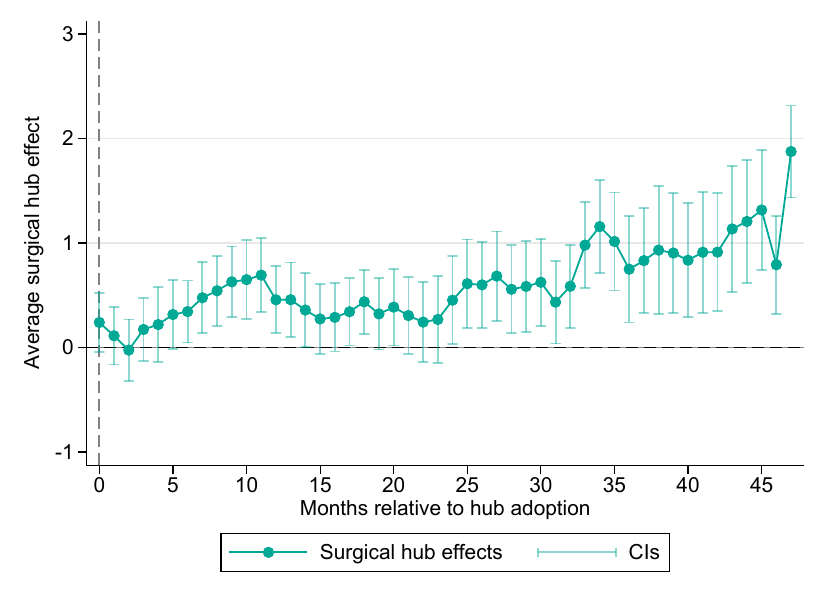}
    \caption{Surgical hub effect over all post treatment periods: Post-adoption event study estimates for the full available horizon (\(k=0\) to \(k=47\)). The effective sample comprises all $\geq 30$ treated units up to \(k=10\), declining thereafter as late adopters exit the post‑treatment window. Estimates beyond \(k=12\) are individually significant and directionally consistent with the main results but are given less interpretive weight given the declining sample. All specifications include hospital fixed effects, month fixed effects, and the full case-mix control vector. Error bars represent 95\% confidence intervals with standard errors clustered at the hospital level.}
    \label{fig:es_long0}
  \end{figure}

\section{Pre-trend test: All pre-treatment periods} 

\begin{figure}[H]
   \centering
    \includegraphics[width=0.8\linewidth]{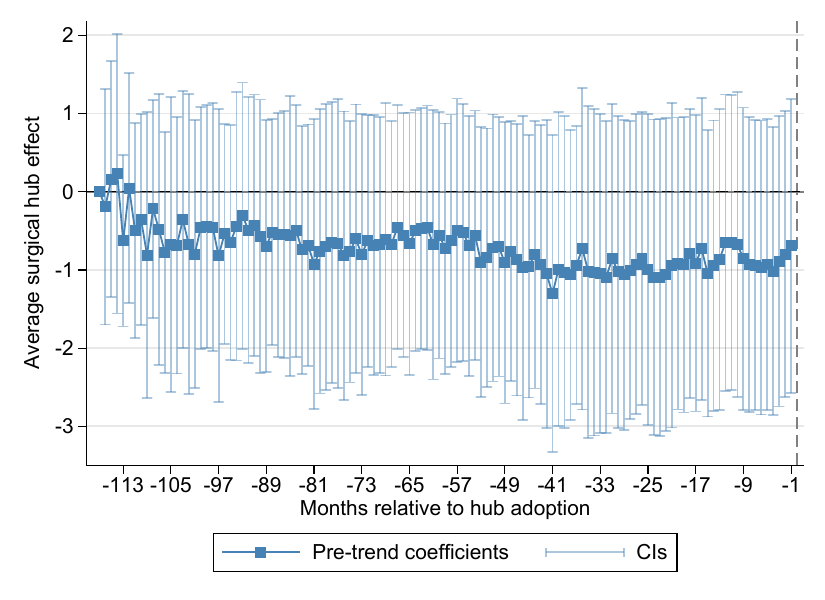}
    \caption{Parallel trends test across all pre-treatment periods. Points are pre-trend lead coefficients $\widehat{\delta}_k$ estimated on untreated observations, covering the full available pre-adoption window. Error bars represent 95\% confidence intervals with standard errors clustered at the hospital level. All specifications include hospital fixed effects, month fixed effects, and the full case-mix control vector. Individual coefficients are uniformly statistically insignificant, consistent with parallel trends. }
    \label{fig:es_long1}
  \end{figure}

Figure \ref{fig:es_long1} extends the pre-trend test to the full available pre-treatment window. Individual pre-treatment coefficients are statistically insignificant across all horizons, consistent with parallel trends. The joint $\chi^2$ test over this extended window returns $p < 0.01$, which we interpret with caution: \citet{borusyak_revisiting_2024} note that joint tests over long pre-treatment windows can over-reject in finite samples as the number of estimated coefficients approaches the effective degrees of freedom, and in our setting the extended window includes COVID-disruption periods for some cohorts, introducing transitory variation that the joint test accumulates but that month fixed effects in the main model absorb. The visual evidence, coefficients with substantially overlapping confidence intervals across all pre-treatment horizons, is more informative than the joint test statistic in this context and supports the parallel trends assumption.

\section{Elective Cancellations: Descriptive Event Study}
\label{app:cancellations}

\begin{figure}[H]
\centering
\begin{subfigure}{0.48\textwidth}
    \centering
    \includegraphics[width=\textwidth]{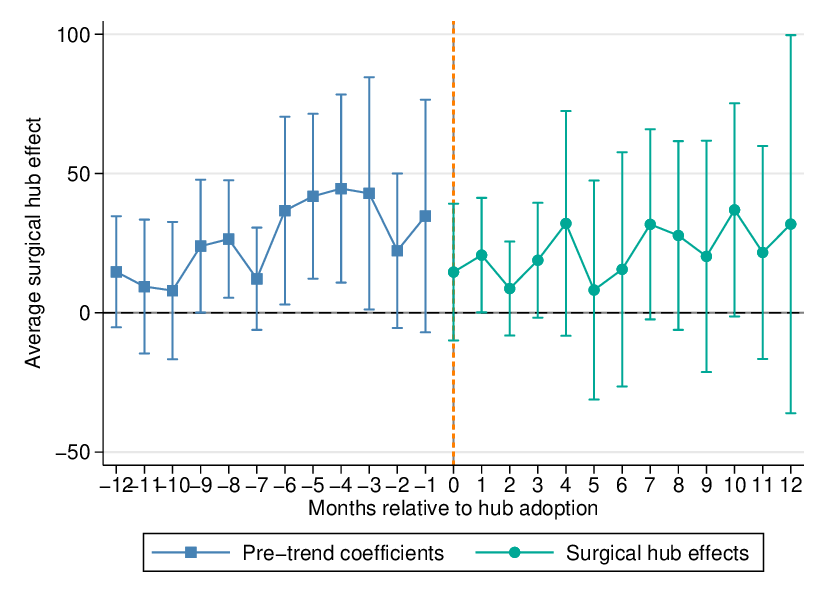}
    \caption{Last-minute elective cancellations}
\end{subfigure}
\hfill
\begin{subfigure}{0.48\textwidth}
    \centering
    \includegraphics[width=\textwidth]{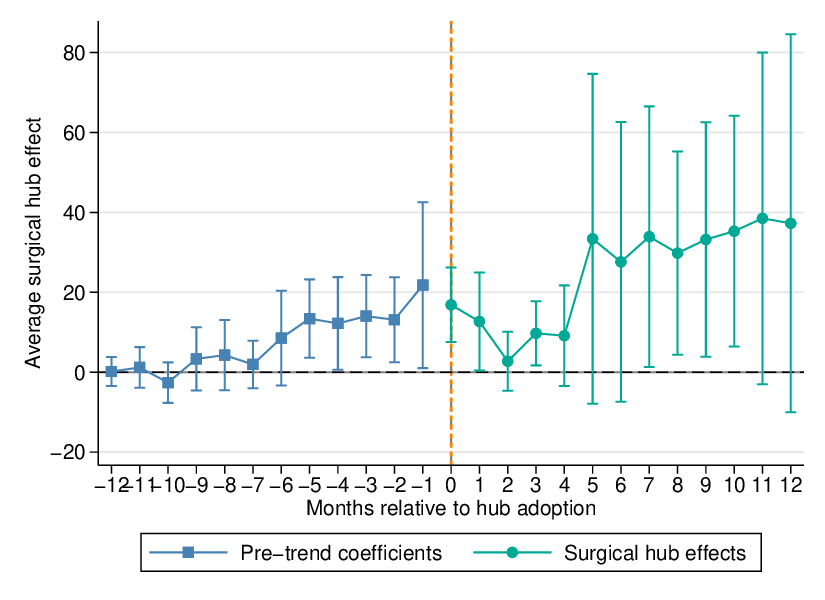}
    \caption{Patients untreated within 28 days 
    post-cancellation}
\end{subfigure}
\caption{Descriptive event study plots for cancellation outcomes. 
Standard errors clustered at the hospital level. All 
specifications include hospital fixed effects, month fixed 
effects, and the full case-mix covariate vector.}
\label{fig:cancellations}
\end{figure}

Figure \ref{fig:cancellations} presents event study plots for 
two cancellation outcomes — last-minute elective cancellations 
and the number of patients untreated within 28 days following 
a cancellation — using quarterly hospital-level data from NHS 
England. These outcomes capture the disruption channel through 
which hub adoption may reduce scheduling interference from 
emergency demand. Pre-trend coefficients show significant 
departures from zero across the pre-adoption window, precluding 
causal inference. Data were not collected during financial year 
2020/21, further limiting the identifying variation. The figures 
should therefore be interpreted as descriptive only.

\section{Robustness Checks}
\label{app:robustness}

\subsection{Alternative Estimators}

Table~\ref{tab:robustness} benchmarks the main BJS estimate against \citet{callaway_difference--differences_2021} (CS) and TWFE across the full post-treatment window and a restricted window of $k = 0$ to $k = 12$ months, over which most treated units contribute to every estimate.

In the full window, the CS estimate ($0.183$, insignificant) is lower and less precise than the BJS baseline ($0.468$). The lower precision reflects an efficiency difference: CS anchors each cohort-specific comparison to a single pre-treatment baseline period, leaving the long pre-treatment history unused, whereas BJS exploits the full pre-treatment window in estimating each hospital's counterfactual \citep{roth_whats_2023}. The lower point estimate is consistent with the two estimators averaging over different sets of cohort-horizon cells over the full window, since the longest horizons are populated only by the earliest-adopting cohorts. Leave-one-hospital-out estimates reported in the Supplementary Material (Section~\ref{S-sec:S4}) indicate that the main result is not disproportionately driven by any single hospital or adoption cohort. The full-window TWFE estimate ($0.425$, signifcant) is close to the BJS baseline, consistent with limited treatment effect heterogeneity bias in this setting.

In the restricted window, BJS yields $0.349$ and TWFE yields $0.341$, both statistically significant. The CS estimate ($0.347$) is of similar magnitude but remains insignificant. The agreement in point estimates across all three estimators is consistent with the long-run parallel trends condition under which BJS is most efficient \citep{dechaisemartin_two-way_2023}.

\begin{table}[!ht]
\centering
\small
\caption{Robustness to Alternative Estimators}
\label{tab:robustness}
\begin{tabular}{lcccc}
\hline
 & \multicolumn{2}{c}{{\ul Full window}} 
 & \multicolumn{2}{c}{\begin{tabular}[c]{@{}c@{}}
   {\ul Restricted window} \\ {\ul ($k = 0$--$12$)}
   \end{tabular}} \\
 & ATT & SE & ATT & SE \\ \hline
BJS  & 0.468$^{***}$ & (0.162) & 0.349$^{**}$ & (0.144) \\
CS   & 0.183         & (0.175) & 0.347        & (0.216) \\
TWFE & 0.425$^{**}$  & (0.190) & 0.341$^{**}$ & (0.169) \\ \hline
\multicolumn{5}{p{8cm}}{\footnotesize \textit{Notes:} 
Outcome: cost-weighted HVLC output per salary-weighted 
physician FTE, deflated to 2019/20 prices. All 
specifications include hospital fixed effects, month fixed 
effects, and the full case-mix covariate vector. CS standard 
errors use the influence-function-based bootstrap; TWFE and 
BJS standard errors are clustered at the hospital level. 
$^{*}p<0.10$, $^{**}p<0.05$, $^{***}p<0.01$.} \\
\end{tabular}
\end{table}

\subsection{Identification Robustness}

Table~\ref{tab:robustness_checks} reports four robustness checks 
using the preferred productivity specification.

\textbf{Region-by-COVID-wave interactions (Column~1).} COVID-19 
waves may have affected hospital productivity differently across 
regions; if hub adoption was geographically concentrated, this 
could confound the treatment estimate. We augment the main model 
with interactions between nine NHS England regional dummies and 
indicators for the four principal COVID-19 waves.\footnote{NHS 
England organises services across nine regions: North East and 
Yorkshire, North West, Midlands, East of England, London, South 
East, South West, East Midlands, and West Midlands. Wave 
classifications follow the UK Health Security Agency's 
epidemiological definitions.} The resulting ATT is 0.446, remaining statistically significant and virtually unchanged from the baseline estimate of $0.468$, suggesting that differential regional COVID exposure does not drive the observed productivity gain.

\textbf{COVID window exclusion (Column~2).} We drop all 
hospital-month observations from April 2020 to March 2021, 
removing the acute pandemic phase from the identifying variation 
and addressing the concern that differential elective recovery 
rather than hub adoption drives the result. The ATT remains stable at $0.455$, close in magnitude to the baseline estimate ($0.468$), leaving the substantive conclusion unchanged.

\textbf{Opening-month exclusion (Columns~3 and~4).} Administrative 
recording changes at hub opening may introduce a mechanical 
component to the $k=0$ productivity jump. Excluding the opening 
month ($k=0$) and then the first two post-adoption months ($k=0,1$) yields ATTs of $0.476$ and $0.491$, marginally larger than the baseline, indicating the opening months are not driving the result.

\begin{table}[!ht]
\centering
\small
\caption{Identification Robustness Checks}
\label{tab:robustness_checks}
\resizebox{\columnwidth}{!}{%
\begin{tabular}{lcccc}
\hline
 & (1) & (2) & (3) & (4) \\
 & Region$\times$Wave & Exclude COVID window 
 & Exclude $k=0$ & Exclude $k=0,1$ \\
\hline \\
$\widehat{\tau}^{\mathrm{ATT}}$ 
  & 0.446$^{***}$ & 0.455$^{**}$ 
  & 0.476$^{***}$ & 0.491$^{***}$ \\
  & (0.159) & (0.179) 
  & (0.164) & (0.166) \\[6pt]
\hline
Region$\times$Wave FE & \checkmark & & & \\
Case-mix covariates 
  & \checkmark & \checkmark & \checkmark & \checkmark \\
Hospital FE   
  & \checkmark & \checkmark & \checkmark & \checkmark \\
Month FE      
  & \checkmark & \checkmark & \checkmark & \checkmark \\
Observations  
  & 11{,}705 & 10{,}446 & 11{,}666 & 11{,}626 \\
\hline
\multicolumn{5}{p{16cm}}{\footnotesize \textit{Notes:} 
Outcome: cost-weighted HVLC output per salary-weighted 
physician FTE, deflated to 2019/20 prices. Standard errors 
clustered at the hospital level in parentheses. 
$^{*}p<0.10$, $^{**}p<0.05$, $^{***}p<0.01$. BJS estimator.} \\
\end{tabular}%
}
\end{table}

\section{Alternative Input Allocation Rules}
\label{app:allocation}

Each specification below modifies the input denominator of the 
preferred productivity measure while holding the output numerator 
unchanged, except E1, which also uses raw HVLC counts as output. 
Two input concepts are used throughout. Total specialty FTE, 
$Z_{ht}$ as defined in Section~\ref{sec:framework}, from 
the Electronic Staff Record covers all contracted physicians in 
the specialty. HVLC physician headcount, denoted $N_{ht}^{\text{HVLC}}$, 
is the count of unique physicians identified in HES APC as 
performing at least one qualifying HVLC procedure at hospital $h$ 
in period $t$.

\textbf{E1 --- HVLC count per HVLC physician headcount.} Input: 
$N_{ht}^{\text{HVLC}}$, the count of unique HVLC physicians 
identified from the HES APC treating consultant identifier. 
Output: $Q_{ht}^{\text{HVLC}}$, the raw (non-cost-weighted) HVLC 
procedure count. This requires no apportionment or salary 
adjustment and serves as a transparent lower-bound benchmark.

\textbf{E2 --- CW-HVLC per HVLC physician headcount.} Identical 
to E1 but with cost-weighted deflated output, $X_{ht}^{\text{HVLC}}$. 
Headcount treats part-time and full-time physicians equally and 
carries no salary weight, whereas the preferred measure uses 
salary-weighted FTE. A hospital with two half-time physicians has 
headcount of two but FTE of one; headcount therefore overstates 
available physician time in such cases.

\textbf{E3 --- CW-HVLC with non-cost-weighted apportionment share.}

\begin{equation}
\phi_{ht}^{E3} = \frac{Q_{ht}^{\text{HVLC}}}{Q_{ht}^{\text{spec}}},
\label{eq:phi_e3}
\end{equation}

\noindent where $Q_{ht}^{\text{spec}}$ is analogously the raw total 
specialty procedure count. Using raw procedure counts rather than 
cost-weighted activity in the share ensures the input denominator 
is independent of the cost weights in the output numerator, 
breaking the potential algebraic link between the two sides of the 
productivity ratio.

\textbf{E4 --- Unapportioned salary-weighted specialty FTE.}

\begin{equation}
Z_{ht}^{E4} = Z_{ht} \cdot w_{ht} / D_t^{\text{pay}}.
\label{eq:L_e4}
\end{equation}

The most conservative specification: attributing all specialty 
physician time to HVLC productivity overstates the input and 
scales down the ratio. Any hub effect surviving this denominator 
cannot be attributed to the apportionment procedure.

\textbf{E5 --- CW-HVLC per HVLC headcount weighted by average FTE.}

\begin{equation}
Z_{ht}^{E5} = N_{ht}^{\text{HVLC}} \cdot 
\frac{Z_{ht}}{N_{ht}^{\text{ALL}}},
\label{eq:L_e5}
\end{equation}

where $N_{ht}^{\text{ALL}}$ is total physician headcount in the 
specialty. This constructs an HVLC-specific input from observed 
physician counts rather than an activity share, without relying on 
the cost-share formula.

\textbf{E6 --- CW-HVLC with pre-treatment apportionment share.}

\begin{equation}
\phi_{h}^{E6} = \bar{\phi}_{h,\text{pre}},
\label{eq:phi_b6}
\end{equation}

averaged over pre-adoption months for treated hospitals and 
pre-2020 months for never-treated hospitals. Fixing the share 
before treatment ensures the input denominator is not affected by 
post-adoption changes in HVLC activity caused by the hub itself.

Table~\ref{tab:robustness_input} reports ATT estimates under each 
specification. The hub effect is positive and significant across 
all six. The unapportioned FTE specification (E4, Column~4) yields the smallest estimate (0.198***), mechanically reflecting the larger denominator. Headcount specifications (E1--E2, Columns~1--2) yield the largest estimates (0.634***), since headcount ignores contracted hours and part-time working. Specifications E3 and E5 (Columns~3 and~5), which break the algebraic link between output and input by using non-cost-weighted and headcount-based shares respectively, yield significant estimates of 0.623** and 0.544***. 
E6 (Column~6), which fixes the share at its pre-treatment mean to 
remove post-adoption simultaneity, yields 0.241***, smaller than 
the baseline but significant. 

The hub effect is robust in sign and statistical significance across all six specifications, though its magnitude varies considerably with the input allocation rule, from 0.198 under the most conservative (unapportioned FTE) denominator to 0.634 under headcount-based measures that ignore part-time working. This range is consistent with the mechanical effect of each denominator choice, rather than genuine instability in the underlying result, and the preferred specification (Section~\ref{sec:framework}) sits toward the middle of this range.

\begin{table}[!ht]
\centering
\caption{Robustness to Alternative Input Allocation Rules}
\label{tab:robustness_input}
\resizebox{\columnwidth}{!}{%
\begin{tabular}{l@{\hspace{8pt}}c@{\hspace{8pt}}c@{\hspace{8pt}}c@{\hspace{8pt}}c@{\hspace{8pt}}c@{\hspace{8pt}}c}
\hline
 & (1) & (2) & (3) & (4) & (5) & (6) \\
 & \begin{tabular}[c]{@{}c@{}}HVLC count\\per HVLC\\headcount\end{tabular}
 & \begin{tabular}[c]{@{}c@{}}CW-HVLC\\per HVLC\\headcount\end{tabular}
 & \begin{tabular}[c]{@{}c@{}}CW-HVLC\\non-CW\\share\end{tabular}
 & \begin{tabular}[c]{@{}c@{}}Unapportioned\\wage-weighted\\FTE\end{tabular}
 & \begin{tabular}[c]{@{}c@{}}CW-HVLC\\headcount\\FTE share\end{tabular}
 & \begin{tabular}[c]{@{}c@{}}CW-HVLC\\pre-treatment\\share\end{tabular} \\
\hline \\
$\widehat{\tau}^{\mathrm{ATT}}$
  & 0.634$^{***}$ & \makebox[2cm][c]{3{,}964.7$^{***}$} & 0.623$^{**}$
  & 0.198$^{***}$ & 0.544$^{***}$ & 0.241$^{***}$ \\
  & (0.194) & \makebox[2cm][c]{(854.4)} & (0.249)
  & (0.058) & (0.179) & (0.078) \\[6pt]
\hline
Case-mix covariates
  & \checkmark & \checkmark & \checkmark
  & \checkmark & \checkmark & \checkmark \\
Hospital FE
  & \checkmark & \checkmark & \checkmark
  & \checkmark & \checkmark & \checkmark \\
Month FE
  & \checkmark & \checkmark & \checkmark
  & \checkmark & \checkmark & \checkmark \\
Observations
  & 11{,}705 & 11{,}705 & 11{,}705
  & 11{,}705 & 11{,}705 & 11{,}705 \\
\hline
\multicolumn{7}{p{19cm}}{\footnotesize \textit{Notes:}
Specifications E1--E6 are described above. Column~(1) also 
uses raw HVLC procedure counts as output. Standard errors 
clustered at the hospital level in parentheses. 
$^{*}p<0.10$, $^{**}p<0.05$, $^{***}p<0.01$. BJS estimator.} \\
\end{tabular}%
}
\end{table}

\subsection{Robustness to Cost Adjustment}

Table \ref{tab:robustness_cost} reports two checks that assess whether the estimated ATT is sensitive to how HVLC output is priced and deflated. 

\textbf{No output deflation (Column~1).} To test whether the real-terms adjustment of output materially affects the estimated ATT, we re-estimate the preferred specification without applying the NHS Cost Inflation Index deflator to the output, retaining year-specific HRG unit costs in nominal terms. The ATT is 0.527, somewhat larger than the baseline of 0.468, which may indicate that without the inflation adjustment the estimated effect would be slightly overstated and that deflating output to 2019/20 prices is appropriate.

\textbf{Frozen pre-2020 HRG unit costs (Column 2).} Our preferred productivity outcome deflates cost-weighted output by an aggregate NHS price index, which removes common inflation but not procedure-specific changes in HRG unit costs that may deviate from the aggregate. NHS England updates its national reference cost schedule annually, meaning if individual procedures performed in hubs were systematically repriced upward after 2020, cost-weighted output could rise post-adoption independently of genuine productivity gains. Column (2) addresses this by replacing annual HRG unit costs with their pre hub programme averages computed over the five financial years from 2014/15 to 2018/19, applied uniformly across the full panel. This eliminates both aggregate inflation and procedure-specific repricing from the output measure. 

The estimated ATT is 0.426 ($p<0.01$), which is close to but somewhat below the baseline estimate of 0.468. Because the only component that differs between the two specifications is the cost weight, the 0.09‑unit reduction, around 9\% of the baseline ATT, is an estimate of the extent to which post‑2020 procedure repricing contributes to the baseline effect. Under frozen costs, the estimated hub effect remains positive and significant, indicating that the majority of the baseline estimate reflects changes in productivity rather than changes in the value of output.

\begin{table}[!ht]
\centering
\caption{Robustness to Cost Adjustment}
\label{tab:robustness_cost}
\resizebox{0.5\textwidth}{!}{%
\begin{tabular}{lcc}
\hline
 & (1) & (2) \\
 & \begin{tabular}[c]{@{}c@{}}No output\\deflation\end{tabular}
 & \begin{tabular}[c]{@{}c@{}}Frozen pre-2020\\HRG unit costs\end{tabular} \\
\hline \\
$\widehat{\tau}^{\mathrm{ATT}}$
  & 0.527$^{**}$ & 0.426$^{***}$ \\
  & (0.223) & (0.145) \\[6pt]
\hline
Case-mix covariates & \checkmark & \checkmark \\
Hospital FE & \checkmark & \checkmark \\
Month FE & \checkmark & \checkmark \\
Observations & 11{,}705 & 11{,}705 \\
\hline
\multicolumn{3}{p{9cm}}{\footnotesize \textit{Notes:}
Column~(1) removes the NHS Cost Inflation Index deflator 
from the output. Column~(2) replaces annual HRG unit costs 
with averages from 2014/15 to 2018/19, applied uniformly 
across the full panel. Input denominator unchanged in both 
columns. Standard errors clustered at the hospital level. 
$^{*}p<0.10$, $^{**}p<0.05$, $^{***}p<0.01$.} \\
\end{tabular}%
}
\end{table}

\end{appendices}

\end{document}


\maketitle

\noindent This supplement reports additional identification, inference, and robustness analyses referenced in the main text. Unless noted otherwise, all specifications use the estimation sample of the preferred baseline specification giving 11,705 hospital-months observations. All include case-mix controls together with hospital and month fixed effects, and cluster standard errors at the hospital level.

\section{Selection into Treatment and Adoption Timing}
\label{sec:S1}

The identification strategy requires that adoption timing is not correlated with the untreated productivity trajectory hospitals would have followed in the absence of hubs. If hospitals that adopt earlier were already on different productivity paths, the imputed counterfactual would be systematically wrong and the estimated effect would reflect those pre-existing differences rather than the programme. To assess this, we examine whether pre-programme characteristics predict adoption and its timing.

We ask two questions. First, do the hospitals that ever adopted a hub differ from those that never did, in characteristics measured before the programme began? Table~\ref{tab:adoption_timing}, column (1), addresses this with a linear probability model for whether a hospital ever adopted. Second, among the 40 adopting hospitals, do those that adopted earlier differ from those that adopted later? Column (2) addresses this with a proportional hazards model for time to adoption, which asks whether a characteristic is associated with adopting sooner rather than later. Both use hospital averages over the pre-programme period from April 2014 to March 2020, together with a hospital-specific trend in productivity over that period, all standardised so that coefficients can be compared across predictors.

The most important predictor is the pre-period productivity trend, since a hospital already on a rising path is precisely the case that would generate a spurious result. Neither model finds any relationship between adoption and the pre-period trend. Nor is there a relationship with the pre-period productivity level in the first model, and the two are jointly insignificant, with a p-value
of 0.276. Larger hospitals were more likely to adopt, which is consistent with the estate and capacity required to establish a hub.

Among adopters, timing shows a weak relationship with the pre-period
productivity level. The hazard ratio of 0.294 lies below one, meaning that hospitals with higher baseline productivity adopted later rather than earlier, so hubs were not directed towards hospitals that were already performing well.
The joint test on level and trend is marginal at 0.094. A permanent difference in productivity level is absorbed by hospital fixed effects and is therefore not a threat to the design. The quantity that would threaten the parallel trends assumption is the pre-period trend, which is insignificant in both models.\footnote{The level result is nonetheless worth stating. If adoption followed a temporary dip in productivity, the subsequent return to normal levels would be misread as a programme effect. The absence of any relationship between adoption timing and the pre-period trend, together with the event study evidence in the main text, argues against this reading.}

\begin{table}[!ht]
\centering
\caption{Predictors of Hub Adoption and Adoption Timing}
\label{tab:adoption_timing}
\IfFileExists{tab_adoption_timing.tex}{\begin{tabular}{lcc}\toprule
            &\multicolumn{1}{c}{Ever adopts (LMP)}&\multicolumn{1}{c}{Time-to-adoption (COX)}\\
 &(1)&(2)\\
\midrule
Pre-period productivity (std.)&       0.091         &       0.294\sym{**} \\
            &     (0.081)         &     (0.175)         \\
\addlinespace
Pre-period productivity trend (std.)&       0.069         &       1.591         \\
            &     (0.053)         &     (0.504)         \\
\addlinespace
Pre-period physician FTE (std.)&       0.226\sym{**} &       1.176         \\
            &     (0.093)         &     (1.136)         \\
\addlinespace
Pre-period HVLC volume (std.)&       0.036         &       1.135         \\
            &     (0.104)         &     (1.004)         \\
\addlinespace
Pre-period day case share (std.)&      -0.053         &       0.911         \\
            &     (0.095)         &     (0.531)         \\
\midrule
p: productivity level \& trend jointly 0&       0.276         &       0.094         \\
Hospitals   &         101         &          40         \\
\bottomrule\multicolumn{3}{p{15cm}}{\footnotesize \textit{Notes:} Hospital level cross-section. column (2) reports hazard ratios; HR$>$1=earlier adoption. Region fixed Effects and pre-period case-mix means included. Robust SEs. Key test: pre-period productivity level and trend should not predict adoption or its timing.} \\ \end{tabular}
}{%
  \begin{tabular}{c}\textit{[Table pending: tab\_adoption\_timing.tex not yet generated]}\end{tabular}%
}
\end{table}

\section{Sensitivity to Violations of Parallel Trends}
\label{sec:S2}

The estimator assumes that, absent the programme, treated and untreated hospitals would have followed parallel productivity paths. 
We show that individual pre-treatment coefficients are statistically insignificant throughout. However, the joint test across the full pre-treatment window rejects the null of no pre-trends at the 1\% level.

Two considerations follow. Joint tests over long windows involve many restrictions and can reject too often in finite samples, so the rejection may overstate the problem \citep{borusyak_revisiting_2024}. Working the other way, a pre-test that fails to reject does not establish that parallel trends holds. With 40 treated hospitals such tests may have limited power, so small but economically meaningful violations could go undetected \citep{roth2022pretest}. Neither argument settles the matter, and resting on either would be unsatisfactory.

We therefore adopt a honest DiD approach, following \citet{rambachan2023more}. Rather than asking whether parallel trends holds, we ask how badly it would have to fail before our conclusion changed. The method allows the assumption to be violated after treatment by some amount, and recomputes the confidence interval under that allowance. The size of the permitted violation is expressed relative to the largest violation actually observed before treatment, and denoted $M$. Setting $M=1$ permits a post-treatment violation exactly as large as the worst pre-treatment deviation present in our own data. Setting $M=2$ permits one twice as large. As $M$ increases the confidence interval widens, and the value at which it first includes zero, the breakdown value, tells us how large a violation the result can withstand.

\begin{figure}[!ht]
\centering
\IfFileExists{fig_honestdid_bjs.pdf}{\includegraphics[width=0.85\textwidth]{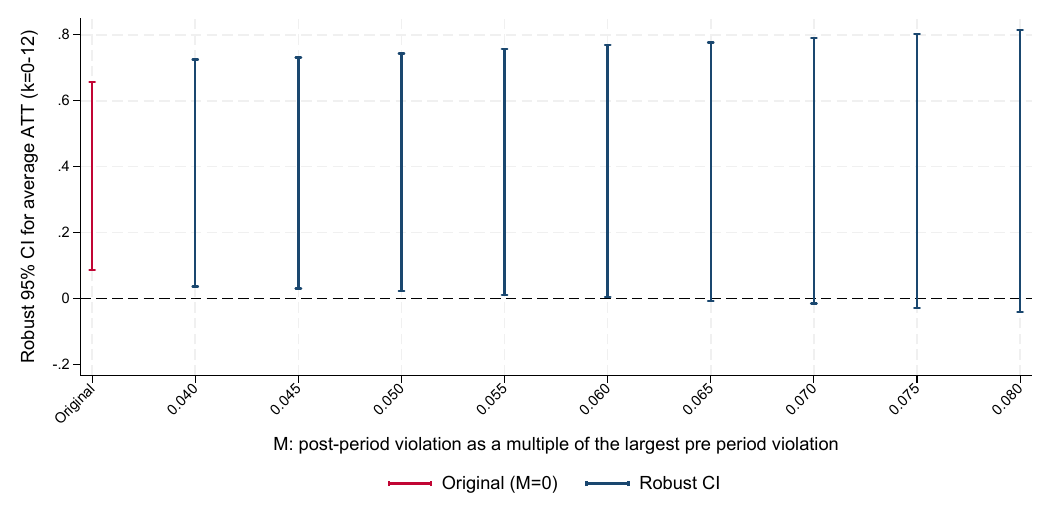}}{%
  \fbox{\parbox{0.85\textwidth}{\centering\vspace{2cm}\textit{[Figure pending: fig\_honestdid\_bjs.pdf not yet generated]}\vspace{2cm}}}%
}
\caption{Robust confidence sets under the relative magnitudes restriction. The horizontal axis gives $M$, the permitted post-treatment violation of parallel trends expressed as a multiple of the largest violation observed before treatment. ``Original'' is the unperturbed confidence interval for the average effect over the first twelve months after adoption.}
\label{fig:honestdid}
\end{figure}

Figure~\ref{fig:honestdid} reports this exercise. The breakdown value is approximately $M=0.06$. The confidence interval retains a positive lower bound at $M=0.060$ and includes zero by $M=0.065$. Violations up to roughly 6\% of the largest pre-treatment deviation therefore leave the main result and conclusion unchanged, while larger departures would overturn it. The assumption under scrutiny is the one conventionally maintained in this literature, that trends found to be parallel before treatment remain so afterwards.\footnote{The analysis uses the restricted post-treatment window covering the adoption month and the twelve months following it, rather than the full window underlying the headline estimate. Beyond twelve months an increasing share of the identification comes from hospitals that adopted in 2020, and at the longest horizons only the earliest cohort contributes, so a full-window analysis would describe the robustness of a progressively narrower group of hospitals. In addition, the permitted deviation accumulates across post-treatment periods under this method, so applying it over 48 horizons would confound the length of the window with robustness. The restricted window is the more favourable of the two choices and we do not claim the headline estimate is more robust than the estimate reported here.}

A different approach to the same concern is to relax the assumption rather than bound the consequences of its failure. We employ a synthetic difference-in-differences \citep{arkhangelsky2021synthetic, clarke2024synthetic}, which gives control hospitals weights chosen so that the weighted control group tracks the pre-treatment path of the treated hospitals, so hospitals whose past behaviour resembles the treated group count for more, and the comparison is made against this reweighted group rather than the raw one. Column 1 of Table~\ref{tab:sdid} reports the result, which is 0.402 somehow smaller in magnitude than the baseline and significant at the 10\% level. 

Standard errors for the imputation estimator are computed from deviations of individual treatment effects around group averages, where the default grouping is by cohort and period. When cohorts are small, each observation contributes substantially to the average it is measured against, so the deviations are mechanically compressed and the standard error is biased downward. \citet{borusyak_revisiting_2024} recommend two adjustments in such settings, computing the averages excluding the own cluster and pooling into coarser groups. Column (2) applies both, using \texttt{leaveout} with \texttt{avgeffectsby(D\_post)}, where \texttt{D\_post} flags post-treatment observations. The standard error rises from 0.162 (on the main estimate) to 0.201, so the main estimate (0.468) remains significant at the 5\% rather than the 1\% level.

\begin{table}[!ht]
\centering
\caption{Synthetic Difference-in-Differences and Alternative Variance Estimator}
\label{tab:sdid}

\begin{tabular}{lcc}\toprule
 & Synthetic DiD & Leave-out SE \\
 & (1) & (2) \\
\midrule
$\widehat{\tau}^{ATT}$ & 0.402\sym{*} & 0.468\sym{**} \\
                       & (0.218)      & (0.201)       \\
                       &  &  \\
\midrule
Observations & 10,080 & 11,705 \\
\bottomrule
\multicolumn{3}{p{12cm}}{\footnotesize \textit{Notes.} Column (1) reports synthetic difference-in-differences on 500 replications, estimated on the balanced subpanel. Column (2) reports the preferred specification from the main text with the alternative variance estimator of \citet{borusyak_revisiting_2024}. The point estimate is identical to the main text by construction and the standard error is larger. Stars denote * $p<0.10$, ** $p<0.05$, *** $p<0.01$.}\\
\end{tabular}
\end{table}

\clearpage

\section{Heterogeneity, Subgroup Composition and Tests of Equality}
\label{sec:S3}

The main text reports separate estimates by hub type, by whether the hub serves a single specialty or several, and by the number of hubs a hospital operates. Where estimates are ordered, as they are across hub types, it is tempting to read the ordering as evidence that deeper separation produces larger gains. Two questions need answering before such a reading is justified. Whether the
subgroups are distinct enough to be interpreted separately, and whether the differences between them are larger than would be expected from sampling variation alone.

Table~\ref{tab:het_crosstab} answers the first by cross-tabulating the 40 treated hospitals across all three dimensions. The dimensions overlap considerably. Operating multiple hubs is concentrated among integrated hubs, five of the eight such hospitals, and single-specialty hubs are a minority throughout, ten of the 38 hospitals for which the classification is recorded. Only eight hospitals operate two or more hubs, so that comparison rests on a small number of hospitals.

\begin{table}[!ht]
\centering
\caption{Cross-Tabulation of Treated Hospitals across Heterogeneity Dimensions}
\label{tab:het_crosstab}
\IfFileExists{tab_het_crosstab.tex}{\begin{tabular}{lcccc}\toprule
 & \multicolumn{2}{c}{Specialisation} & \multicolumn{2}{c}{Number of hubs}\\
Hub type & Single & Multi & One & Two+ \\\midrule
Integrated & 2 & 10 & 7 & 5 \\
Ring-fenced & 3 & 9 & 11 & 1 \\
Standalone & 5 & 9 & 14 & 2 \\
\bottomrule\multicolumn{5}{p{9cm}}{\footnotesize \textit{Notes:} Counts of the 40 treated hospitals. Hub type refers to the first hub opened. Specialisation is unrecorded for two standalone hospitals, so the specification columns sum to 38 rather than 40.}\\
\end{tabular}}{%
  \begin{tabular}{c}\textit{[Table pending: tab\_het\_crosstab.tex not yet generated]}\end{tabular}%
}
\end{table}

Table~\ref{tab:het_tests} answers the second. A difference between two estimates is not evidence of a real difference unless it is large relative to the uncertainty in each. We therefore test equality directly. Because conventional standard errors can understate uncertainty when a subgroup contains few hospitals, we report a wild cluster bootstrap alongside the conventional test. The bootstrap does not rely on the same large-sample approximation and is more reliable when the number of clusters is small. We also report the equality test computed within the imputation estimator itself, which is the estimator used for the point estimates in the main text.

Three conclusions follow. Differences by hub type are not statistically distinguishable under any of the three approaches, with a joint p-value of 0.928 under two-way fixed effects and 0.862 under the imputation estimator, and a standalone against ring-fenced comparison giving 0.716 and 0.594. The ordering of point estimates across hub types is consistent with a separation gradient but does not establish one, and we describe it accordingly in the main text. The specialisation comparison depends on which estimator is used, being insignificant under two-way fixed effects at 0.396, with a bootstrap value of 0.437, but significant under the imputation estimator at 0.014. We treat it as unresolved. The comparison by number of hubs is significant under all three approaches, at 0.003, 0.004 and 0.000, and is the one heterogeneity result we regard as robust. It carries the caveat that hospitals operating several hubs are also larger, so the comparison may partly reflect hospital scale rather than the number of hubs as such.

\begin{table}[!ht]
\centering
\caption{Tests of Equality of Subgroup Effects}
\label{tab:het_tests}
\IfFileExists{tab_het_tests.tex}{\begin{tabular}{p{2.6cm}p{5.2cm}ccc}\toprule
 & & TWFE & TWFE & BJS \\
Panel & Contrast & Wald & wild boot. & hetby/lincom \\
 & & (1) & (2) & (3) \\
\midrule
Hub type & Standalone = Integrated = Ring-fenced (joint) & 0.928 & --- & 0.862 \\
         & Standalone vs Ring-fenced & --- & 0.716 & 0.594 \\
Specialisation & Single vs Multi-specialty & 0.396 & 0.437 & 0.014 \\
Number of hubs & One vs Two+ & 0.003 & 0.004 & 0.000 \\
\bottomrule
\multicolumn{5}{p{15cm}}{\footnotesize \textit{Notes:} All entries are p-values for the null that the subgroup effects being compared are equal. Columns (1) and (2) come from interacted two-way fixed effects specifications with case-mix controls, the first using a conventional clustered Wald test and the second a wild cluster bootstrap (Rademacher weights, 9,999 replications). Column (3) tests the same contrast within the imputation estimator used for the point estimates in the main text, using its own joint variance-covariance matrix. We report all three because they need not agree. The analytic clustered standard error can understate uncertainty when a subgroup contains few hospitals, only eight in the case of multi-hub adoption, which is why the bootstrap is included as a size robust check. Where columns (2) and (3) disagree we treat the conclusion for that contrast as estimator sensitive rather than settled.}\\
\end{tabular}}{%
  \begin{tabular}{c}\textit{[Table pending: tab\_het\_tests.tex not yet generated]}\end{tabular}%
}
\end{table}

\section{Influence Diagnostics}
\label{sec:S4}

With 40 treated hospitals adopting a hub across a limited number of cohorts, the estimate may depend on a small number of hospitals or on a single adoption cohort rather than reflecting the programme as a whole. A related concern arises from the outcome itself. Productivity is measured as a ratio, and ratios are sensitive to small values in the denominator, since these can produce large values that exert disproportionate influence on the estimate.

\begin{figure}[!ht]
\centering
\IfFileExists{fig_loo.pdf}{\includegraphics[width=\textwidth]{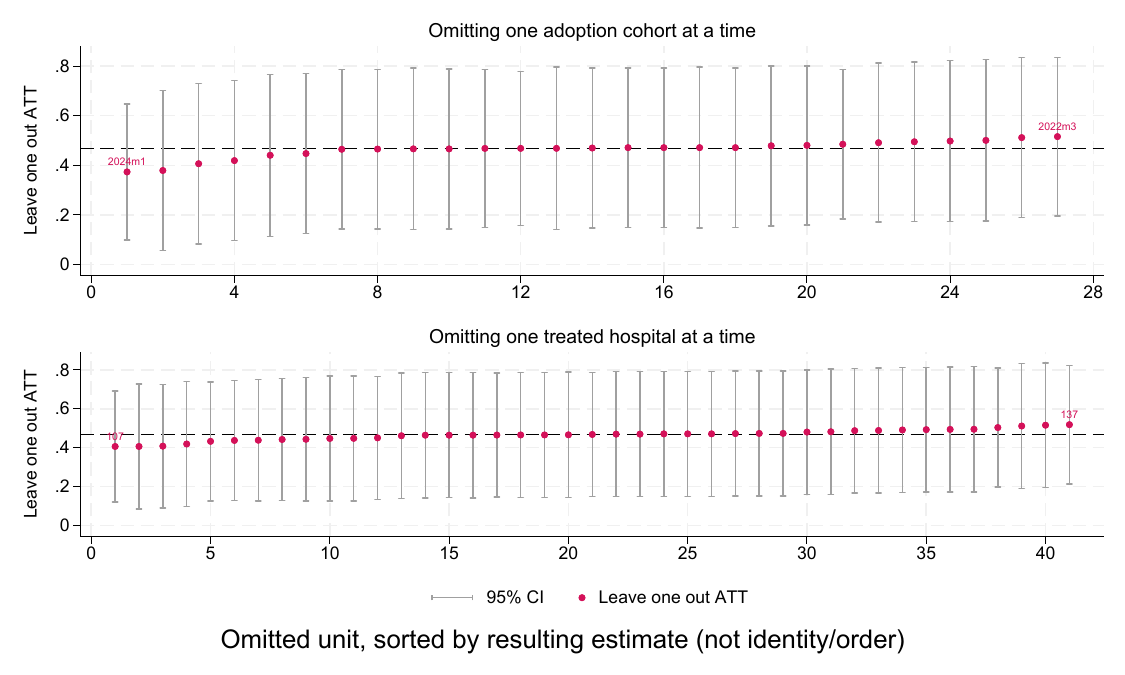}}{%
  \fbox{\parbox{\textwidth}{\centering\vspace{2cm}\textit{[Figure pending: fig\_loo.pdf not yet generated]}\vspace{2cm}}}%
}
\caption{Leave-one-out estimates. The top panel omits one adoption cohort at a time and the bottom panel omits one treated hospital at a time. Units are sorted by the resulting estimate rather than by identity or adoption order, so the upward slope carries no meaning beyond the ordering itself. The dashed line is the baseline estimate, and extreme points are labelled by adoption month or hospital identifier.}
\label{fig:loo}
\end{figure}

Figure~\ref{fig:loo} addresses the first. It re-estimates the baseline specification once for each adoption cohort, omitting that cohort, shown in the top panel. The lower panel repeats the exercise omitting one treated hospital at a time. If the result depended on a particular hospital or cohort, removing it would move the estimate substantially. It does not. All estimates lie between approximately 0.38 and 0.52, and every confidence interval contains the baseline estimate. The largest movement comes from omitting the most recently adopting cohort, January 2024, which contributes the fewest post-treatment observations. Excluding it yields approximately 0.38, which remains statistically significant.

Table~\ref{tab:ri} addresses the second. Excluding the top and bottom 1\% of the outcome distribution yields 0.396. Excluding hospital-months in the bottom percentile of apportioned physician time yields 0.450. Both remain statistically significant. The modest attenuation indicates that extreme observations exert some influence without accounting for the result.

\begin{table}[!ht]
\centering
\caption{Sensitivity to Extreme Observations}
\label{tab:ri}
\IfFileExists{tab_ri.tex}{\begin{tabular}{lcc}\toprule
 & ATT & (SE) \\
 & (1) & (2) \\\midrule
Baseline & 0.468\sym{***} & (0.162) \\
Trimmed outcome (1st--99th pct.) & 0.396\sym{***} &  (0.138) \\
Excl. near zero input months & 0.450\sym{***} &  (0.156) \\
\bottomrule\multicolumn{3}{p{11cm}}{\footnotesize \textit{Notes:} Observations outside the trimming range are dropped rather than replaced with the boundary value, so that the estimate is computed on the remaining sample without imposing a value on the excluded hospital-months. * $p<0.10$, ** $p<0.05$, *** $p<0.01$}\\
\end{tabular}

}{%
  \begin{tabular}{c}\textit{[Table pending: tab\_ri.tex not yet generated]}\end{tabular}%
}
\end{table}

\clearpage 

\section{Robustness to Industrial Action}
\label{sec:S5}

Elective activity in England was disrupted by junior doctor industrial action between January 2023 and February 2024 \citep{garner2026evaluating}. This matters for our design in two ways. Hospitals that adopted late have post-treatment windows falling largely within this period, so their estimated effects are measured against disrupted activity. And because exposure to disruption varied across hospitals, treated and untreated hospitals may have been affected differently.\footnote{Strike dates are taken from \href{https://medichut.com/the-uk-junior-doctor-strikes-of-2023-and-2024/}{this summary of the 2023 and 2024 junior doctor strikes},}

Table~\ref{tab:strikes} reports three tests. The first simply removes the months in which industrial action took place, so that no comparison is drawn from a strike month. The estimate is essentially unchanged at 0.467 against a baseline of 0.468. The second removes the entire period from January 2023 to February 2024, including the months between strikes, on the reasoning that lists deferred ahead of announced action and activity catching up afterwards would fall in those months. This yields 0.427. Part of the difference reflects the loss of most of the post-adoption window for the latest adopting cohorts rather than industrial action as such, since the panel ends in March 2024.

The third test takes a different approach. Rather than removing months, it asks whether hospitals more vulnerable to disruption behaved differently during strike months. We measure vulnerability by each hospital's rate of last minute non-clinical elective cancellations before the pandemic, which is a measure of how readily its elective pathway is disturbed. This is determined well before both hub adoption and the strikes, so it cannot itself be an outcome of either. Interacting it with strike months leaves the estimate unchanged at 0.469, and the interaction is small and statistically insignificant.\footnote{This measures general fragility of the elective pathway rather than exposure to junior doctor withdrawal specifically, and is taken more than three years before the strikes, so it should be read as an attenuated test. The month exclusion results are the primary evidence.} Excluding the affected months leaves the estimate essentially unchanged, which indicates that industrial action does not account for the estimated productivity gain.

\begin{table}[!ht]
\centering
\caption{Robustness to the 2023 to 2024 Industrial Action}
\label{tab:strikes}
\IfFileExists{tab_strikes.tex}{\begin{tabular}{l*{4}{c}}\toprule
            &\multicolumn{1}{c}{Baseline}&\multicolumn{1}{c}{Excl. strike months}&\multicolumn{1}{c}{Excl. Jan23-Feb2024}&\multicolumn{1}{c}{Cancellation Exposure}\\
            & (1)  & (2) &  (3) &  (4) \\
\midrule
$\widehat{\tau}^{ATT}$&       0.468\sym{***}&       0.467\sym{***}&       0.427\sym{**} &       0.469\sym{***}\\
            &     (0.162)         &     (0.163)         &     (0.166)         &     (0.162)         \\
 &   &  &   &   \\
 \midrule
           
Observations&      11,705         &      10,937         &      10,361         &      11,705         \\
\bottomrule\multicolumn{5}{p{15cm}}{\footnotesize \textit{Notes:} Strike months are those with junior doctors and/or consultants industrial action affecting NHS elective care in England (January, March--June and August 2023; January--February 2024). Column (3) additionally excludes the intervening months of the industrial action period. Column (4) interacts strike months with each hospital standardised pre pandemic rate of last minute elective cancellations. SEs clustered at hospital level.}\\ \end{tabular}
}{%
  \begin{tabular}{c}\textit{[Table pending: tab\_strikes.tex not yet generated]}\end{tabular}%
}
\end{table}

\bibliographystyle{agsm}
\bibliography{references}